\documentclass[10pt]{article}

\usepackage{amsmath,amssymb,amsthm,bm,graphicx,cite,url,natbib}

\title{Quantitative system risk assessment from incomplete data with belief networks and pairwise comparison elicitation}
\author{Cristina De Persis\thanks{Avio S.p.A., Italy}, 
Jos\'{e} Luis Bosque\thanks{Department of Computer Science and Electronics, Universidad de Cantabria, Spain}, 
Irene Huertas\thanks{ESA ESTEC, Keplerlaan 1, 2201 AZ Noordwijk, Netherlands}, 
Simon P.~Wilson\thanks{School of Computer Science and Statistics, Trinity College Dublin, Ireland}}
\date{}

\begin{document}
\maketitle

\begin{abstract}
A method for conducting Bayesian elicitation and learning in risk assessment is presented. It assumes that the risk process can be described as a fault tree. This is viewed as a belief network, for which prior distributions on primary event probabilities are elicited by means of a pairwise comparison approach.  A Bayesian updating procedure, following observation of some or all of the events in the fault tree, is described.  The application is illustrated through the motivating example of risk assessment of spacecraft explosion during controlled re-entry.

\end{abstract}

\section{Introduction}
The belief network, otherwise known as a Bayesian network or directed acyclic graph, is a powerful tool for the probabilistic modelling of systems of random variables. Originally proposed for use in the quantitative risk assessment of systems at least as early as \cite{barlow88}, its relationship to a fault tree and the reliability block diagram was recognized quite quickly \citep{bobbio01, torres98}. The general methodology is well described in papers such as \cite{langseth07} and \cite{neil07}. It provides a structure around which to elicit the system risk in terms of its causal events, update the elicited assessment with data and explore the risk's sensitivity to the probability of those causes.  It permits the modelling of dependent failure modes e.g.\ through a common cause \citep{mi12}. The structure of the belief network can be derived from a fault tree analysis, and indeed it can be viewed as a generalization that allows richer probabilistic relationships between events in the tree.  It is particularly suited to assessments where there are uncertainties in the relationships between events in the system and the risk of interest.
 
In this paper we focus on its use as a generalization of a fault tree analysis, where the main interest is in learning about the probability of the top event, and where a fault tree has been constructed that relates it to the occurrence of one or more primary or intermediate causal events.  Furthermore, we focus on a common situation where:
\begin{enumerate}
\item There is substantial expert opinion but that constraints on the availability of the experts, or their experience of an elicitation process, meaning that the elicitation process must be kept simple;
\item There are limited data from past instances of the risky event, meaning that this information must be used to the full but that also there is uncertainty in the risk assessment that must be properly quantified.  In other words, typically some of the events in the fault tree are not observed, and which are observed may change from one observation to the next.
\end{enumerate}
This situation is ubiquitous when one is discussing risks associated with the emergence of a new technology or where a new risk is identified for an existing system. The benefits of the approach described here are a light elicitation burden on the experts and the proper management of uncertainties through the Bayesian paradigm.


The use of a panel of experts for risk analysis has been studied at some length; see \cite{aspinall13} and \cite{ohagan06} for example. One common approach, and the one we adopt here, is that of elicitation by pairwise comparison that has been used since the early days of elicitation methods \cite{gulliksen59, guo10}. It is recognized as one of the simpler ways of accessing expert opinion. A method motivated by the analytic hierarchy process is described, although equally any other pairwise comparison approach could be used. 

The motivating example for this work comes from an application in the space industry. To reduce the creation of space debris, operators are increasingly resorting to a controlled re-entry of satellites and spacecraft once they are no longer needed, with the objective that they will largely burn up in the atmosphere.   The re-entry trajectory is designed so that any components that do reach the surface will land in areas of remote ocean, such as the South Pacific. A recent example, and the motivation for this work, is the European Space Agency's Automated Transfer Vehicle (ATV), built to supply the International Space Station.  Other examples of situations where the approach of this paper may be relevant are nuclear power \cite{verma15}, maritime safety \cite{vandorp10,zhang16} or counter-terrorism \cite{merrick10}.

This paper begins with a brief description of fault trees, belief networks and their relationship, and sets up notation.  In Section \ref{sec:pairwise}, the use of pairwise comparison to elicit priors on the primary event probabilities is discussed.  Section \ref{sec:data} explores how the elicited risk assessment can be updated with data that gives only partial information about the events in the tree. Section \ref{sec:conc} presents the main conclusions of the work.
Section \ref{sec:ATV} describes the application of the approach to spacecraft re-entry.


\section{Fault trees and belief networks}
\label{sec:set_up}

\begin{figure}
\centering
\includegraphics[scale=0.26]{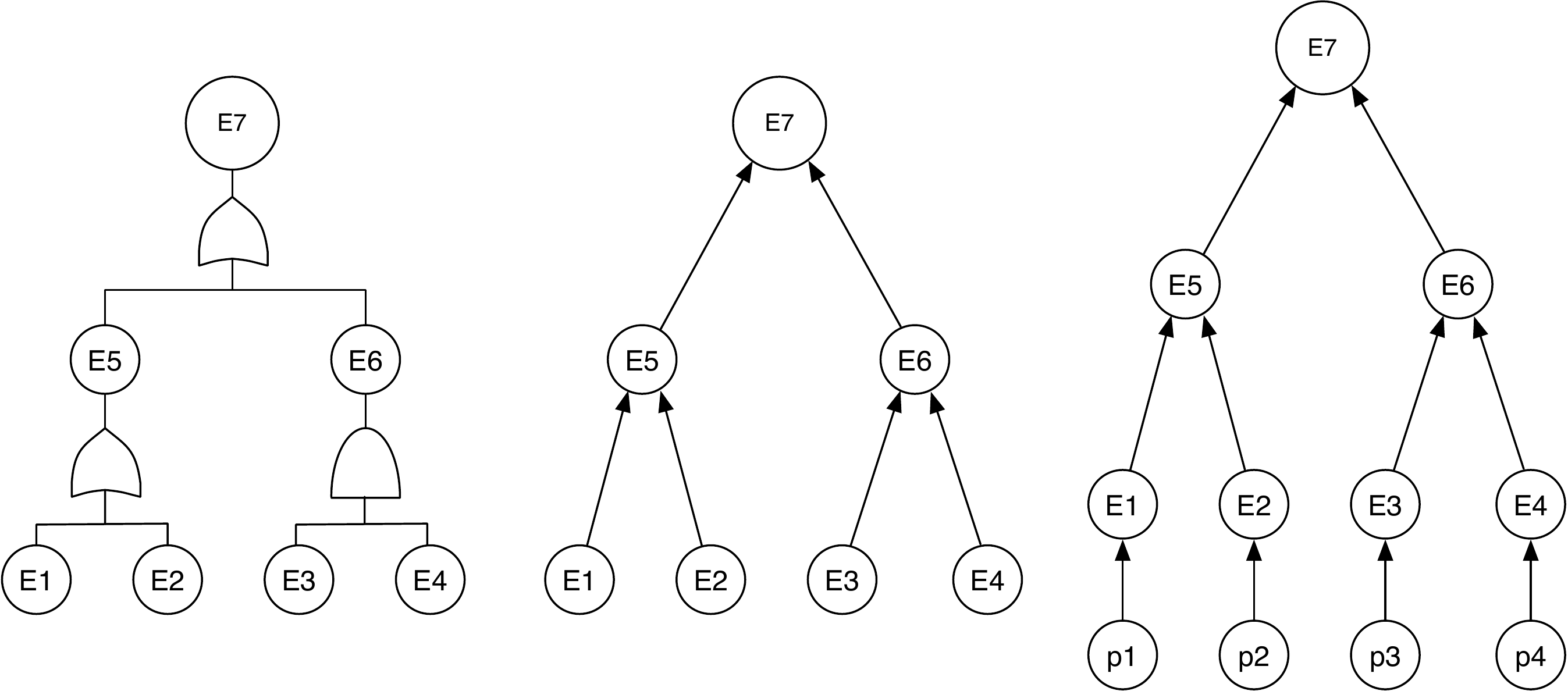}
\caption{\label{fig:simple_example} From left to right: a fault tree for a system with 4 primary events ($E_1$ to $E_4$), 2 intermediate events ($E_5$ and $E_6$) and the top event $E_7$, the equivalent representation as a belief network and the belief network extended to include the primary events probabilities $p_i = P(E_i = 1)$, $i=1,\ldots,4$.}
\end{figure}

Consider a fault tree, such as on the left of Figure \ref{fig:simple_example}, describing the logical relationship between a set of events in a system, culminating in the top event of interest.  The value of each node in the tree is in $\{0,1, \mbox{NA}\}$, with 1 indicating that the event has been observed to occur, 0 indicating that it has been observed not to have occurred, and NA indicating that it has not been observed.  Aside from the top event, and following standard terminology, a primary or base event has no causes developed further in the tree, while an intermediate event is the consequence of other events in the tree. The logic in this tree can be represented as a directed acyclic graph or belief network that brings it into the probabilistic risk modelling domain \cite{bobbio01}. In this representation, primary events are assumed to be independent. This logic can be enriched with probabilistic rather than deterministic relationships between events by specifying probability tables for the occurrence of each event in terms of the values of its parents in the network.

Leaving aside for the moment the issue of whether an event is observed, let there be $n$ events in the tree, with $E_i \in \{0,1\}$ representing whether event $i$ occurred or not, $i=1,\ldots,n$.  Let $E_{1:i} = \{ E_1,\ldots,E_i \}$ represent the first $i$ of these events and let  $\eta_i \subset \{E_1,\ldots,E_n\}$ denote the immediate causal events of event $i$ in the tree, also known as the parents of $E_i$ in the language of belief networks.  If event $i$ is primary then $\eta_i = \emptyset$.  Let $H_i$ represent all ancestors (parents, parents of parents, etc.) of $E_i$. Assume that there are $k < n$ primary events and that they are labelled $E_1,\ldots,E_k$, and also assume that the top event is $E_n$. For the primary events, let $p_i = P(E_i = 1)$ and let $\bm{p} = (p_1,\ldots,p_k)$ denote the set of primary event probabilities. Also note that, because of the tree structure of the network, we can label events in the tree such that the index of all ancestors of $E_i$ are indexed before $i$, by first labelling the primary events, then the children of primary events, etc.\ so that $H_i \subseteq E_{1:i-1}$. It will also be useful to define the successors (children, grandchildren, etc.) of $E_i$ as $S_i$.

\begin{figure}
\centering
\includegraphics[scale=0.4]{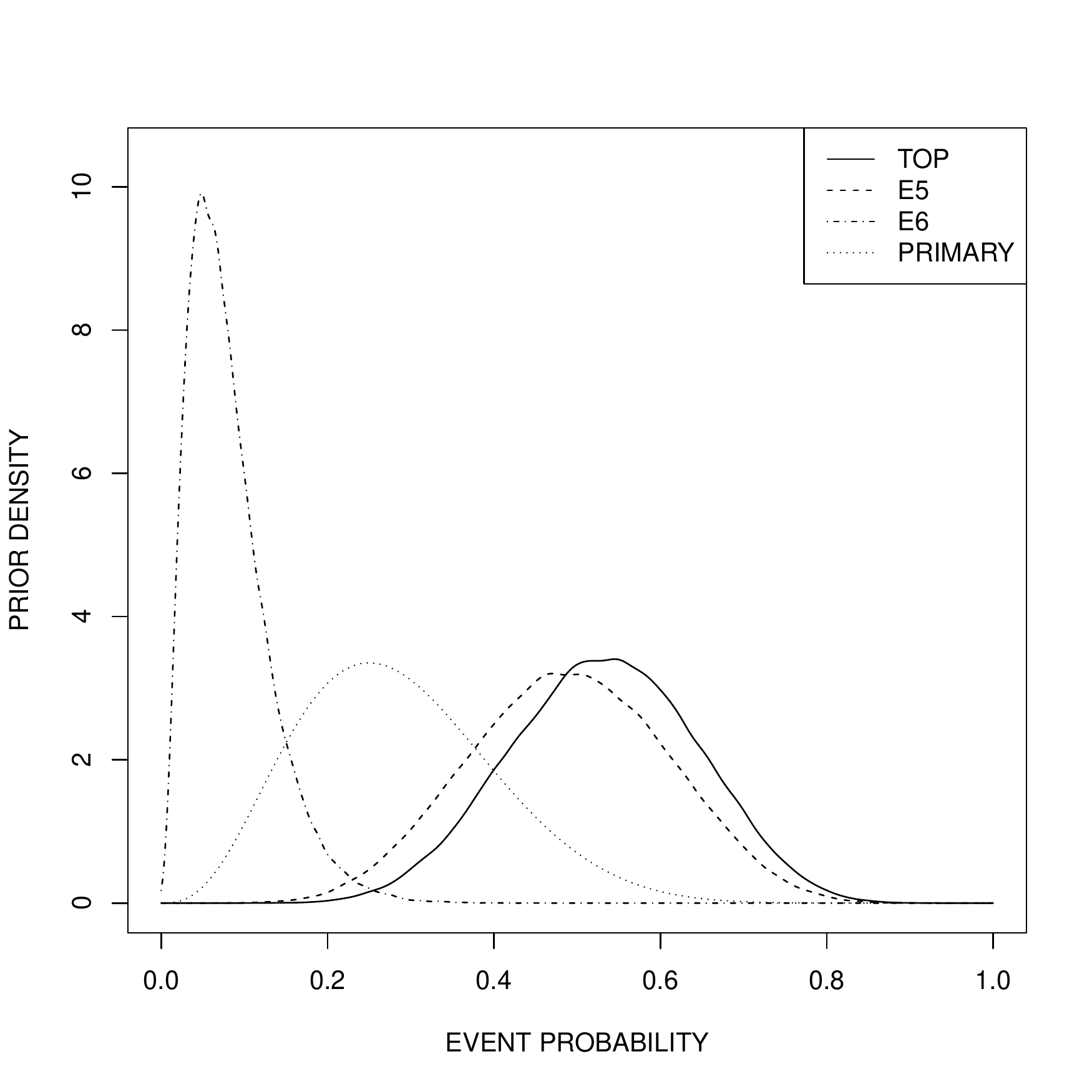}
\caption{\label{fig:prior_simulation} Kernel density estimates of the marginal prior distributions of $p_5$, $p_6$ and the top event probability $p_7$ for the example in Figure \ref{fig:simple_example}, based on samples from a beta(4,10) prior distribution on each of the 4 primary events.}
\end{figure}

In the belief network representation, stochastic nodes for each $p_i$ are added and they are assigned prior distributions, usually beta distributions for reasons of conjugacy. The network logic then implies prior distributions for the probability of intermediate events and the top event. In other words, when the belief network just models the logic of the fault tree, prior specification of the primary event probabilities is sufficient to specify the prior of the probability of any event in the tree. For example, the probability $p$ that an event occurs, in terms of 2 independent causal events that occur with probabilities $p_1$ and $p_2$ is given by $p = 1 - (1-p_1)(1-p_2)$ in the case of an OR gate in the tree, or $p = p_1 \, p_2$ in the case of an AND gate. The prior distribution $f(p)$, in terms of the prior distribution $f_1(p_1,p_2)$, is then obtained by the usual rules of distribution of functions of a random variable \cite[Chapter 6]{grimmett14}. These distributions are often not in a closed form but in practice are approximated by Monte Carlo simulation from the primary event prior. 
For example, assuming independent beta(4,10) prior distributions for each of the 4 primary events in Figure \ref{fig:simple_example}, Figure \ref{fig:prior_simulation} shows kernel density estimates of the marginal prior distributions of $p_5 = 1 - (1-p_1)(1 - p_2)$, $p_6 = p_3 \, p_4$ and the top event probability
\begin{equation*}
p_7 \: = \: 1 - (1-p_5)(1-p_6) \: = \: 1 - (1-p_1)(1-p_2)(1-p_3 p_4)
\end{equation*}
based on 10,000 simulations of $(p_1,p_2,p_3,p_4)$.
The prior distribution of $p_7$ is then the initial risk assessment based on the specification of the prior distributions for the primary events.


%


\section{Primary Event Probability Prior Elicitation}
\label{sec:pairwise}
Following construction of the fault tree, prior distributions for the $p_i$ are elicited from expert opinion. Many elicitation methods are now available \cite{dias2018,garthwaite05}, and any of these can in principle be applied at this stage.  In this work, focus is on constructing independent beta prior distributions for each $p_i$, because of their subsequent tractability.  Since the amount of time that experts can devote to elicitation is often short --- in our spacecraft re-entry example, the experts were engineers with many other demands on their time --- a simpler elicitation scheme, such as one based on pairwise comparisons, is attractive. There is an extensive literature on this approach. \cite{bradley52} describe an early approach, which is discussed and compared in some detail by \cite{cooke91}; \cite{szwed06} is a more recent use. Here we describe an alternative pairwise comparison approach. 

The expert is asked to select the 'cornerstone' primary event $E_{i^*}$ that he or she has most confidence in giving a prior distribution of its probability. A standard elicitation approach is then used to specify a beta prior distribution of this event's probability. In this work, a range of values of the probability $(p^{(L)},p^{(U)})$ is elicited and the prior is specified to be the beta distribution that has this range as its central 95\% probability interval; computing this range is an easy numerical exercise since the quantiles of the beta distribution can be accurately and quickly approximated.  

Then the expert is asked a series of pairwise comparison questions to rate whether each other primary event is more or less likely to occur than the cornerstone event.  The expert is asked to rate the probability of a primary event as being equally, moderately, strongly, very strongly or absolutely more or less than another. At a minimum, each primary event is compared in this way with the cornerstone event.  Ideally all pairwise comparisons would be elicited. Where $E_i$ and $E_j$ are compared, the comparison of $E_j$ to $E_i$ is assumed to be the reverse e.g.\ if $E_i$ is strongly more probable than $E_j$ then $E_j$ is assumed to be strongly less probable than $E_i$. These 9 qualitative comparisons are mapped to a numerical score, motivated by the similar approach in the analytic hierarchy process (AHP) method for eliciting preferences \cite{saaty80}, whose use in prior elicitation has been recognized e.g. \cite{cagno00}. 
The comparison scores are placed in a matrix ${\cal Q} = (q_{ij})$, with $q_{ij}$ representing the score of the comparison between $E_i$ and $E_j$.   These scores are then mapped to a weight $w_{i}$ for each event through the geometric mean approach of \cite{crawford85}, where a smaller weight indicates an event that has been elicited to have a lower probability of occurring (see the Appendix). 

The weights are used to specify a range of values for each $p_i$ as
\begin{equation}
\frac{w_i}{w_{i^*}} p^{(L)} \leq p_i \leq \min\left( 1, \frac{w_i}{w_{i^*}} p^{(U)} \right),
\label{eq:prob_bounds_reweight}
\end{equation}
where $w_i^*$ is the weight of the cornerstone event.  This interval is then mapped to a beta prior distribution that has it as its central 95\% probability interval. This has the effect of shifting prior weight to lower values of $p_i$ when $E_i$ is rated less likely on average than $E_{i^*}$.

\begin{figure}
\begin{minipage}{0.45\textwidth}
\begin{center}
{\scriptsize
\begin{tabular}{lc|c} \hline
Comparison ($E_1$ to $E_2$) & Desired prior  & Score    \\ 
                            & mean on $p_2$           &  \\ \hline
Absolutely less probable    & 0.95               & 0.17 \\
Very strongly less probable & 0.85               & 0.21 \\ 
Strongly less probable      & 0.75               & 0.28 \\
Moderately less probable    & 0.60               & 0.53 \\
Equally probable            & 0.50               & 1.00 \\
Moderately more probable    & 0.40               & 1.04 \\
Strongly more probable      & 0.25               & 1.23 \\
Very strongly more probable & 0.15               & 1.52 \\ 
Absolutely more probable    & 0.05               & 2.55 \\ \hline
\end{tabular}
}
\end{center}
\end{minipage}
\quad\quad\quad
\begin{minipage}{0.45\textwidth}
\begin{center}
\includegraphics[scale=0.35]{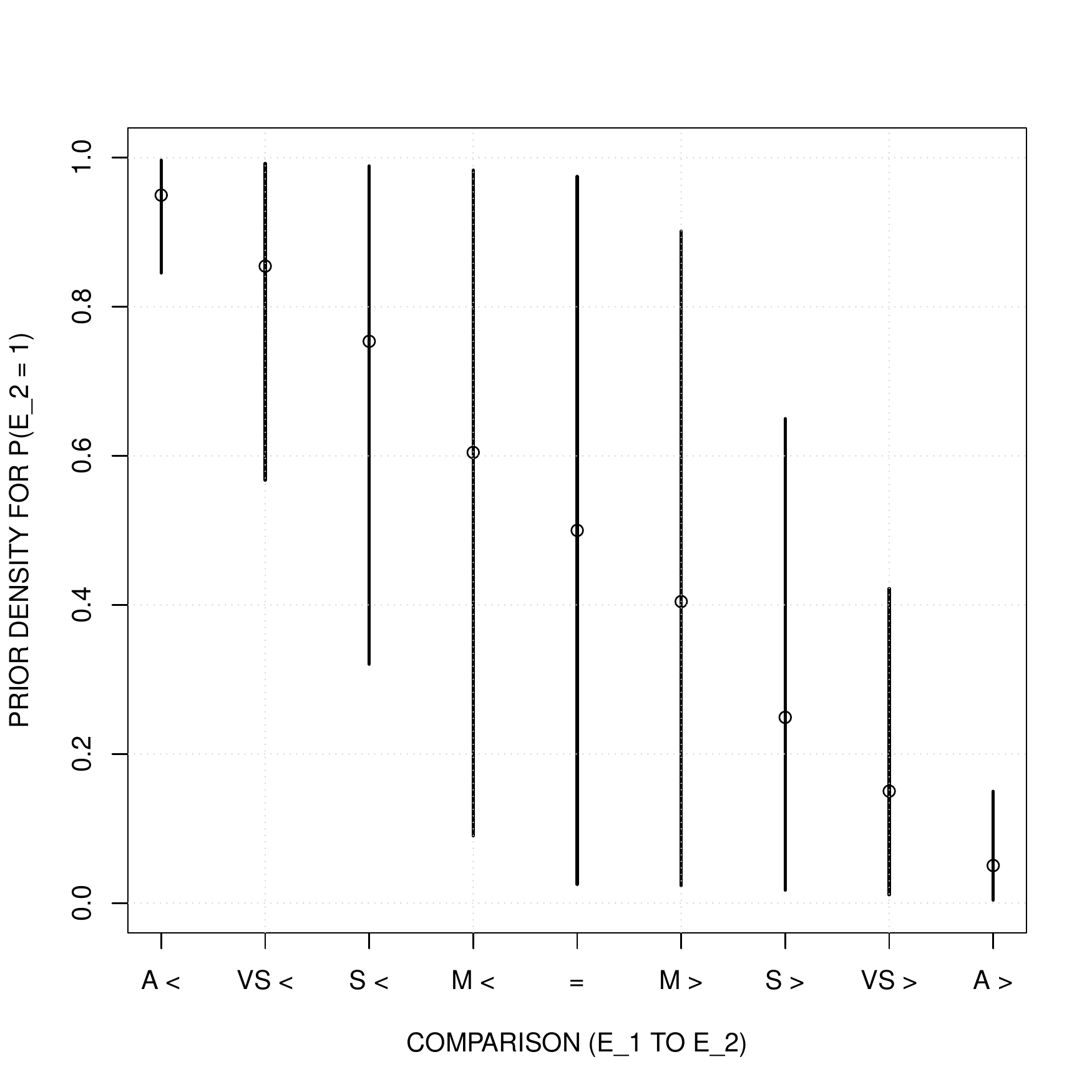}
\end{center}
\end{minipage}
\caption{\label{fig:ahp_scores} The mapping of qualitative pairwise comparisons to a score (left) and the effect of these on the elicited prior for $p_2$ in the case of comparing it with a uniform prior on $p_1$ (right). The circle is the prior mean and the line indicates the central 95\% prior probability.}
\end{figure}

Figure \ref{fig:ahp_scores} shows the scores that we use.  These scores are derived from the following simple example of 2 events, with $E_1$ as the cornerstone event.  Assuming that a uniform prior is assessed on $p_1$, so with mean 0.5, the scores are derived that give a prior mean for $p_2$ from 0.95 to 0.05 (from an absolutely less probable to absolutely more probable comparison) under the above procedure.

\begin{figure}
\parbox{\columnwidth}{
  \parbox{0.45\columnwidth}{
    \centering
     \begin{eqnarray*}
          {\cal Q} & = & \left(
                          \begin{tabular}{cccc}
                          1.00 & 0.21 & 1.04 & 0.53 \\
                          1.52 & 1.00 & 1.04 & 1.52 \\
                          0.53 & 0.53 & 1.00 & 0.17 \\
                          1.04 & 0.21 & 2.55 & 1.00
                          \end{tabular}
                          \right) \\
                   &   & \\          
                 w & = & (0.162, 0.444, 0.120, 0.273) \\
     \mathbb{E}(p_{1:4}) & = & (0.020, 0.068, 0.015, 0.043) \\
     \mathbb{E}(p_7) & = & 0.088
     \end{eqnarray*}
  }
  \parbox{0.45\columnwidth}{
    \centering
     \begin{eqnarray*}
          {\cal Q} & = & \left(
                          \begin{tabular}{cccc}
                          1.00 & 0.21 & 1.04 & 0.53 \\
                          1.52 & 1.00 & --   & --   \\
                          0.53 & --   & 1.00 & --   \\
                          1.04 & --   & --   & 1.00
                          \end{tabular}
                          \right) \\
                   &   & \\          
                 w & = & (0.136, 0.425, 0.148, 0.291) \\
     \mathbb{E}(p_{1:4}) & = & (0.020, 0.077, 0.023, 0.054) \\
     \mathbb{E}(p_7) & = & 0.097
     \end{eqnarray*}
  }}  
\parbox{\columnwidth}{  
  \parbox{0.45\columnwidth}{
    \centering
     \includegraphics[scale=0.4]{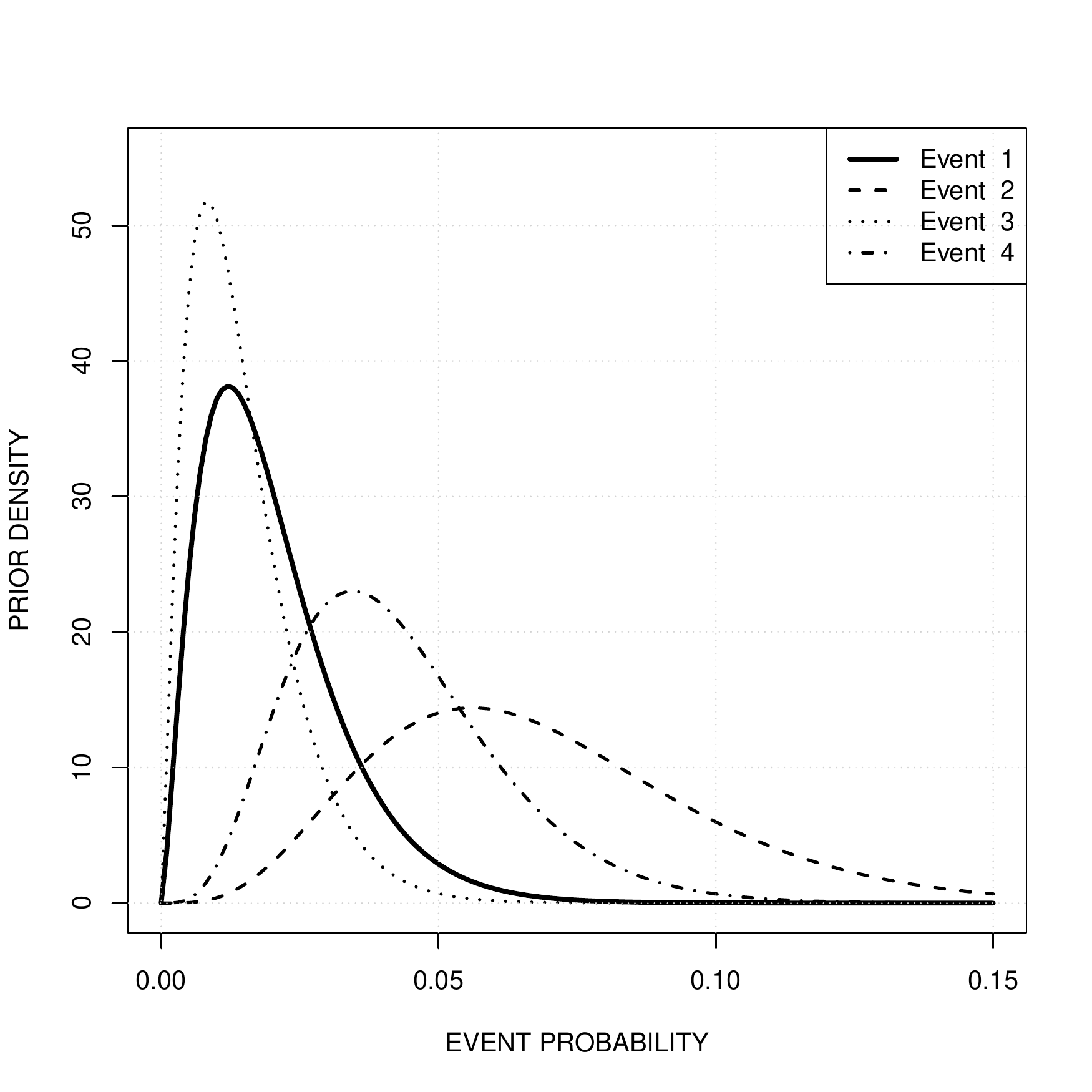}
  }
  \parbox{0.45\columnwidth}{
    \centering
     \includegraphics[scale=0.4]{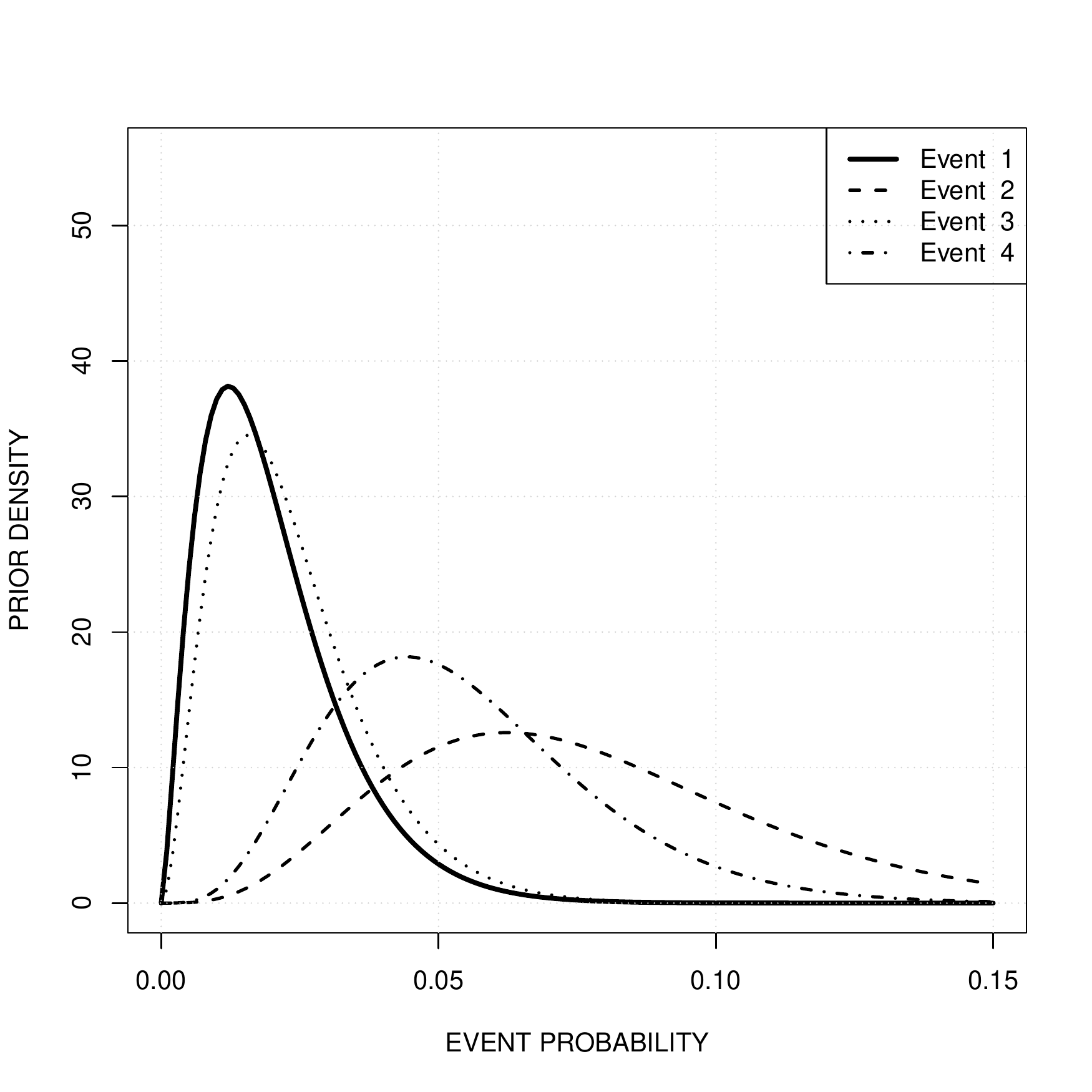}
  }}
\parbox{\columnwidth}{  
  \parbox{0.45\columnwidth}{
    \centering
     \includegraphics[scale=0.4]{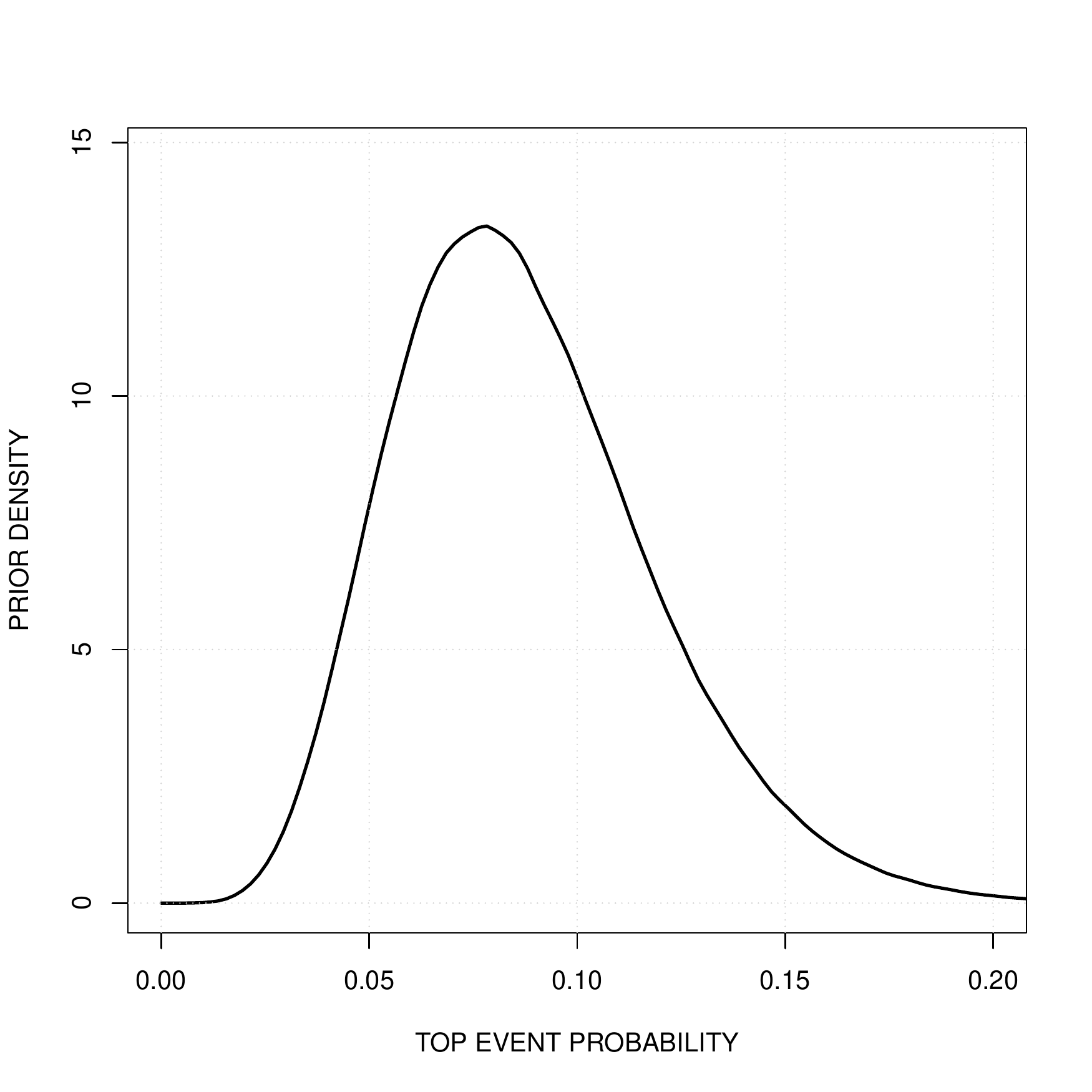}
  }
  \parbox{0.45\columnwidth}{
    \centering
     \includegraphics[scale=0.4]{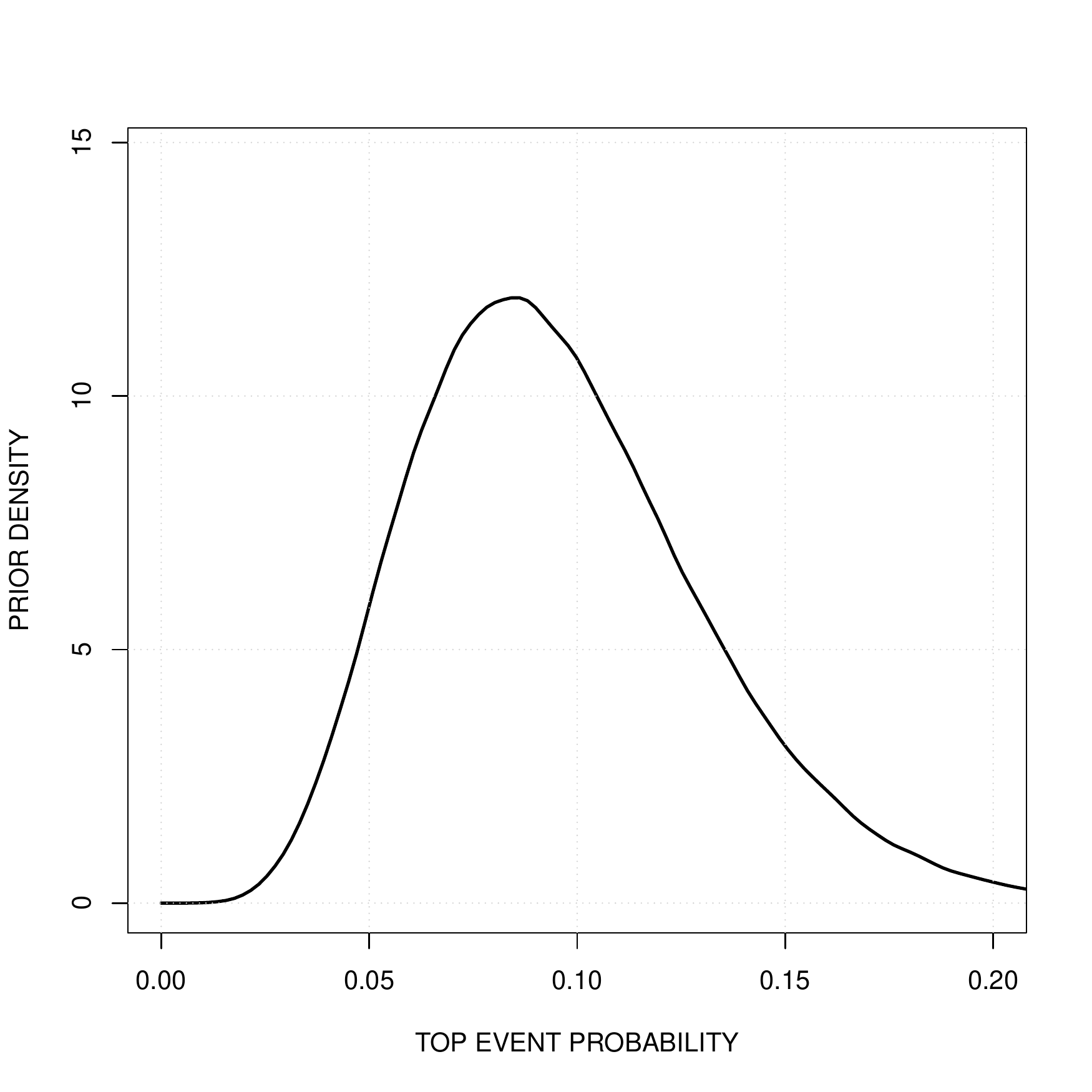}
  }}  
\caption{\label{fig:elicit_example_full} Elicitation of 4 primary events.  Event 1 is the cornerstone event with elicited interval $(0.01,0.05)$.}
\end{figure}

As a simple example, consider the system with 4 primary events as in Figure \ref{fig:simple_example}. Figure \ref{fig:elicit_example_full} illustrates the prior elicitation stage where $E_1$ is deemed to be the cornerstone event and the expert gives an interval $(0.01,0.05)$ as the range for its probability, which equates to a beta distribution with parameters 2.5 and 120.  The left column of the plot shows the case where a full set of pairwise comparisons is made, as given in the matrix ${\cal Q}$, from which the method of \cite{crawford85} derives weights $w$. The figure then shows the resulting beta prior distributions for the primary events, following Equation \ref{eq:prob_bounds_reweight} for the other 3 events as well as event 1 and then finally the resulting prior distribution on the top event probability. The prior expectation of each primary event $p_i$ and for the top event is also given. The right column shows the case where only comparisons with the cornerstone event are available.

\section{Incorporation of Partial Data about the Event}
\label{sec:data}
Once the fault tree is defined and primary event probability prior distributions are elicited, the risk assessment can be updated with data about the event. In this section, it is derived for 2 cases: the complete case, where all $n$ events in the tree are observed, and the incomplete case, where only a subset of the events are observed.

\subsection{Likelihood for a Complete Observation}
\label{subsec:complete}
The likelihood is $P(E_1,\ldots,E_n \, | \, \bm{p})$ in this case. Recalling the standard result of a belief network that the joint distribution of the variables in the network is the product of the probabilities of each variable given its parents, we have
\[ P(E_1,\ldots,E_n \, | \, \bm{p}) = \prod_{i=1}^n P(E_i \, | \, \eta_i, \bm{p}). \]
Since $E_1,\ldots,E_k$ are the primary events that are assumed independent, with $P(E_i \, | \, p_i) = p_i^{E_i} (1-p_i)^{1-E_i}$, one has
\begin{multline}
p(E_1,\ldots,E_n \, | \, \bm{p}) \:
 = \: \left( \prod_{i=1}^k P(E_i \, | \, p_i) \right) \left(\prod_{i=k+1}^n P(E_i \, | \, \eta_i, \bm{p}) \right) \\
= \: \left( \prod_{i=1}^k p_i^{E_i} (1-p_i)^{1-E_i} \right) \left(\prod_{i=k+1}^n P(E_i \, | \, \eta_i, \bm{p}) \right), \\
 = \: \left( \prod_{i=1}^k p_i^{E_i} (1-p_i)^{1-E_i} \right) \left(\prod_{i=k+1}^n P(E_i \, | \, \eta_i) \right);
\label{eq:full_likelihood}
\end{multline}
the last line comes from the independence of non-primary $E_i$ from $\bm{p}$ given $\eta_i$. The value of these $E_i$ are logically derived from the fault tree. If $E_i$ is the result of an AND gate then $E_i=1$ if and only if $E_j=1$ for all $E_j \in \eta_i$, while if it is the result of an OR gate then $E_i=1$ if and only if $E_i=1$ for some $E_j \in \eta_i$. Hence $P(E_i \, | \, \eta_i)$ is either 1 or 0, depending on whether the values of $E_i$ and $\eta_i$ are consistent with the logic of the fault tree or not.
To summarize, the likelihood for $\bm{p}$ from a complete observation, given by Equation \ref{eq:full_likelihood}, is a product of the Bernoulli likelihoods of each primary event as long as the logic of the fault tree is respected for all the non-primary event values.  If it is not respected then the likelihood is 0.

\subsection{Likelihood for an Incomplete Observation}
\label{subsec:incomplete}
The more common scenario in situations that interest this paper is that only a subset ${\cal E} \subset \{E_1,\ldots,E_n\}$ is observed. The likelihood is now $P({\mathcal E} \, | \, \bm{p})$.

A general approach to deriving this likelihood is by marginalisation of the complete likelihood over the events that are not in ${\mathcal E}$:
\begin{eqnarray} 
P({\mathcal E} \, | \, \bm{p}) 
& = & \sum_{\substack{E_i = 0,1\\E_i \notin {\mathcal E}}} P(E_1,\ldots,E_n \, | \, \bm{p}) \nonumber \\
& = & \sum_{\substack{E_i = 0,1\\E_i \notin {\mathcal E}}}  \left( \prod_{i=1}^k p_i^{E_i} (1-p_i)^{1-E_i} \right) \: \times \: \left(\prod_{i=k+1}^n P(E_i \, | \, \eta_i) \right).
\label{eq:incomplete_likelihood}
\end{eqnarray}
Factoring out any observed primary events from the sum gives
\begin{equation}
P({\mathcal E} \, | \, \bm{p}) \: = \: \left( \prod_{\substack{i=1\\E_i \in {\mathcal E}}}^k p_i^{E_i} (1-p_i)^{1-E_i} \right) \: \times \:  \sum_{\substack{E_i = 0,1\\E_i \notin {\mathcal E}}}  \left( \prod_{\substack{i = 1\\E_i \notin {\mathcal E}}}^k p_i^{E_i} (1-p_i)^{1-E_i} \right) \:
\times \: \left(\prod_{\substack{i = k+1\\E_i \notin {\mathcal E}}}^n P(E_i \, | \, \eta_i) \right).
\end{equation}
The final term $\prod_{i=k+1}^n P(E_i \, | \, \eta_i)$ over intermediate events is either 1 or 0, as in Section \ref{subsec:complete}, depending on whether  that combination of observed and unobserved event values in $E_i$ and $\eta_i$ are consistent with the logic of the fault tree or not. Thus the likelihood is a product of the Bernoulli likelihoods for observed \textit{primary} events, multiplied by a sum of Bernoulli likelihoods for all combinations of unobserved primary events such that the fault tree logic is respected. Hence, one way to derive the likelihood in this case is to go through each combination of values of the \textit{unobserved} events, check to see if the fault tree logic is respected, and if it is then add the term $\prod_{\substack{i=1\\E_i \notin {\mathcal E}}}^k p_i^{E_i} (1-p_i)^{1-E_i}$ to a sum.  Once that is done, the sum is multiplied by $\prod_{\substack{i=1\\E_i \in {\mathcal E}}}^k p_i^{E_i} (1-p_i)^{1-E_i}$.

As an example, take the simple system of Figure \ref{fig:simple_example} where we observe $E_1 = 0$, $E_3 = 1$ and  $E_6 = 1$; no other events are observed. $E_1$ and $E_3$ are primary events and contribute $(1-p_1) p_3$ to the likelihood.  There are 2 unobserved primary events, $E_2$ and $E_4$, and 2 of the 4 possible combinations of those are consistent with the observation of $E_6$, namely $(E_2,E_4) = (0,1)$ and $(1,1)$, contributing $(1-p_2) p_4 + p_2 p_4 = p_4$ to the likelihood, for a likelihood of the form $(1-p_1) p_3 p_4$.

This approach to constructing the likelihood permits common cause events, that is an event can be the parent of more than one event.  A disadvantage is that it does not scale to a situation where there are a large number of unobserved events, because of the large number of combinations of them that then must be considered.  

\subsection{Construction of an Incomplete Likelihood in the Pure Tree Case}
When the belief network follows a tree structure, so that each event is the parent of at most one other event, then there is a constructive approach to building the likelihood that may still be practical even when there are a large number of unobserved events. The tree structure constraint means that there cannot be common cause events; each event must be the cause of at most one other event for this approach to work. 

First note that in a tree, all nodes are ''converging'' (e.g.\ 2 or more parents have a unique child) apart from the primary nodes, which have a unique $p_i$ as their parent.  In such a converging node, the parents are independent given that the child is not observed.  Hence one can partition ${\mathcal E}$ into subsets that consist of an event $E_i$ that has no observed successors, and all its observed ancestors $H_i \cap {\mathcal E}$, and these subsets will be independent. Recalling that $S_i$ is the set of successors of $E_i$, we have:
\begin{equation}
P({\mathcal E} \, | \, \bm{p}) \: = \: \prod_{\substack{E_i \in {\mathcal E}\\ S_i \cap {\mathcal E} = \emptyset}} P(E_i, H_i \cap {\mathcal E} \, | \, \bm{p}) \:
 = \: \prod_{\substack{E_i \in {\mathcal E}\\ S_i \cap {\mathcal E} = \emptyset}} P(H_i \cap {\mathcal E} \, | \, \bm{p}) \: P(E_i \, | \, H_i \cap {\mathcal E}, \bm{p}).
\label{eq:incomplete_recursive}
\end{equation}
If the top event $E_n$ is observed then there is no strict partition and Equation \ref{eq:incomplete_recursive} becomes $P({\mathcal E} \, | \, \bm{p}) \: = \: P(H_n \cap {\mathcal E} \, | \, \bm{p}) \: P(E_n \, | \, H_n \cap {\mathcal E}, \bm{p})$.

As regards the 2 terms on the right hand side of Equation \ref{eq:incomplete_recursive}:
\begin{itemize}
\item For $P(E_i \, | \, H_i \cap {\mathcal E}, \bm{p})$, either $E_i$ is logically implied from $H_i \cap {\mathcal E}$, in which case $P(E_i \, | \, H_i \cap {\mathcal E}, \bm{p}) = 1$, or it is not. If it is not then we can write this probability as a function of the unobserved primary ancestor event probabilities of $E_i$:
\begin{equation}
P(E_i \, | \, H_i \cap {\mathcal E}, \bm{p}) \:
= \: g_i(\{ p_j \, | \, 1 \leq j \leq k, \: E_j \in H_i \cap \overline{\mathcal E}\});
\label{eq:incomplete_gi}
\end{equation}
see the Appendix for the derivation of the function $g_i$.
\item One can apply Equation \ref{eq:incomplete_recursive} recursively to $P(H_i \cap {\mathcal E} \, | \, \bm{p})$, partitioning $H_i \cap {\mathcal E}$ into subsets by the events in $H_i \cap {\mathcal E}$ that have no observed successors in $H_i \cap {\mathcal E}$.  The recursion ends when $H_i \cap {\mathcal E} = \emptyset$, in which case $P(E_i, H_i \cap {\mathcal E} \, | \, \bm{p}) = P(E_i \, | \, \bm{p})$ and one follows the derivation in the Appendix.
\end{itemize}
Recursive application of Equations \ref{eq:incomplete_recursive} and \ref{eq:incomplete_gi} will yield the expression for the likelihood $P({\mathcal E} \, | \, \bm{p})$.  Note that this method favours the situation where there are few observed events, in contrast to the marginalisation approach of Section \ref{subsec:incomplete}.
 
\subsection{Posterior Computation}
\label{subsec:compute}
For ease of notation, in this section ${\mathcal E}$ could also refer to a complete observation as well as an incomplete one.  The posterior distribution of $\bm{p}$ given a set of $m$ such observations ${\mathcal E}_1,\ldots,{\mathcal E}_m$ is then 
\[ P(\bm{p} \, | \, {\mathcal E}_1,\ldots,{\mathcal E}_m) \propto P(\bm{p}) \: \prod_{l=1}^m P({\mathcal E}_l \, | \, \bm{p}), \]
where $P({\mathcal E}_l \, | \, \bm{p})$ has been derived using of the methods of the previous parts of this section.

Unfortunately the likelihood is not conjugate to the beta prior distributions on the $p_i$ and so posterior calculation is implemented by Monte Carlo methods.  If $m$ is not too large then importance sampling can be used to generate samples of $\bm{p}$ from the posterior distribution with $P(\bm{p})$ as the proposal distribution: a large sample of values $\bm{p}_1,\ldots,\bm{p}_R$ is generated from the prior (an easy task as it is a product of independent beta distributions), weights $\omega_r = \prod_{l=1}^m P({\mathcal E}_l \, | \, \bm{p}_r)$ are calculated and a posterior sample comes from sampling the $\bm{p}_r$ with probabilities proportional to the $\omega_r$. Alternatively, a random walk Metropolis sampling scheme can be used. This has been done with zero-mean normal proposals on the logit $p_i$; given a current $\bm{p}$, propose $\lambda_i^* \sim N(\lambda_i, s^2_{\lambda})$, where $\lambda_i = \log(p_i) - \log(1-p_i)$, from which the proposal is $p_i^* = e^{\lambda_i^*}/(1+ e^{\lambda_i^*})$. The proposed vector $\bm{p}^*=(p_1^*,\ldots,p_k^*)$ is accepted with probability 
\[ \min \left\{ 1, \frac{p(\bm{p}^*) \: \prod_{l=1}^m p({\mathcal E}_l \, | \, \bm{p}^*)}{p(\bm{p}) \: \prod_{l=1}^m p({\mathcal E}_l \, | \, \bm{p})} \right\}. \]
This approach works better when $m$ is sufficiently large that the prior and posterior are significantly different.

The posterior samples can be used to approximate the posterior distribution of any intermediate event or the top event by further simulation, as described in Section \ref{sec:set_up}.

\subsection{Summary and Example}
A full risk assessment procedure, with calibration from data, has now been described.  The procedure starts with an expert or experts constructing a fault tree with binary events. A pairwise comparison procedure, such as that described in Section \ref{sec:pairwise}, leads to the specification of a prior distribution $P(\bm{p})$ and hence, by the fault tree logic, to a prior on the probability of the top event, which is usually approximated by Monte Carlo simulation. This is the output for the initial risk assessment.  If a point risk estimate is required then the prior median or mean can be used.

This assessment can be updated with data following observation of an instance of the risky situation. The prior distribution of $\bm{p}$ is updated to a posterior distribution that can be evaluated by Monte Carlo methods.  These posterior samples can be used to generate samples from the posterior distribution of the top event probability. Similarly, the median or mean of these samples can be used as a point estimate for the probability.

\begin{figure}
\parbox{\columnwidth}{  
  \parbox{0.3\columnwidth}{
     \textsf{\small{\textbf{Data}}}
  }   
  \parbox{0.35\columnwidth}{
    \centering
    \textsf{\small{\textbf{Primary Event Probabilities}}}
  }
  \parbox{0.35\columnwidth}{
    \centering
    \textsf{\small{\textbf{Top Event Probability}}}
}}
\parbox{\columnwidth}{  
  \parbox{0.3\columnwidth}{
     \textsf{\small{Prior Elicitation\\
     $\mathbb{E}(\bm{p}) \\ = (0.020,0.077,0.023,0.054)$\\
     $\mathbb{E}(p_7)  =  0.097$}}
  }   
  \parbox{0.35\columnwidth}{
    \centering
    \includegraphics[scale=0.25]{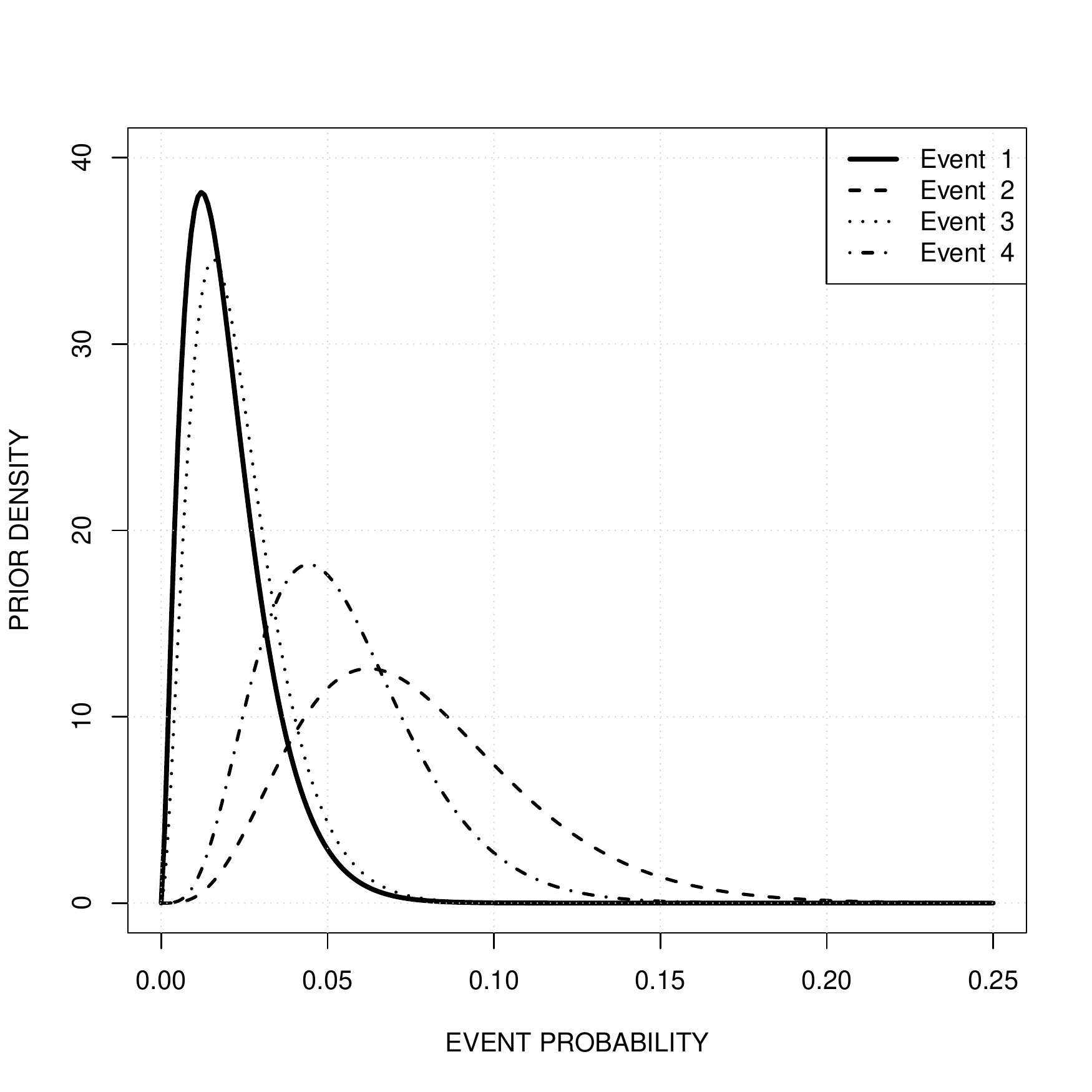}
  }
  \parbox{0.35\columnwidth}{
    \centering
    \includegraphics[scale=0.25]{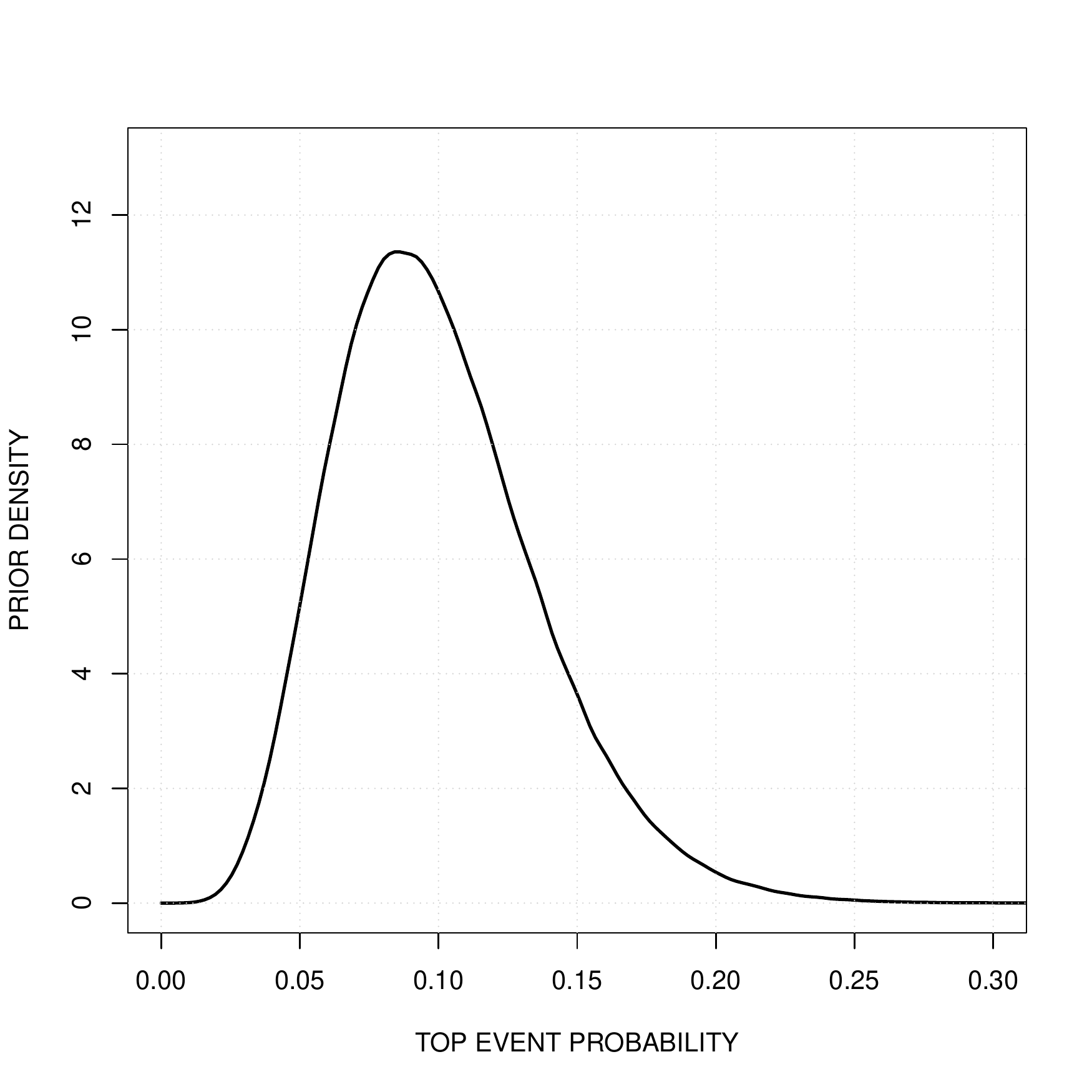}
}} 
\parbox{\columnwidth}{  
  \parbox{0.3\columnwidth}{
     \textsf{\small{Complete data \\
     $\mathbb{E}(\bm{p} \, | \, \mathcal{E}_{1:40}) \\ = (0.010, 0.058, 0.021, 0.050)$ \\
     $\mathbb{E}(p_7 \, | \, \mathcal{E}_{1:40}) = 0.068$}}
  }   
  \parbox{0.35\columnwidth}{
    \centering
    \includegraphics[scale=0.25]{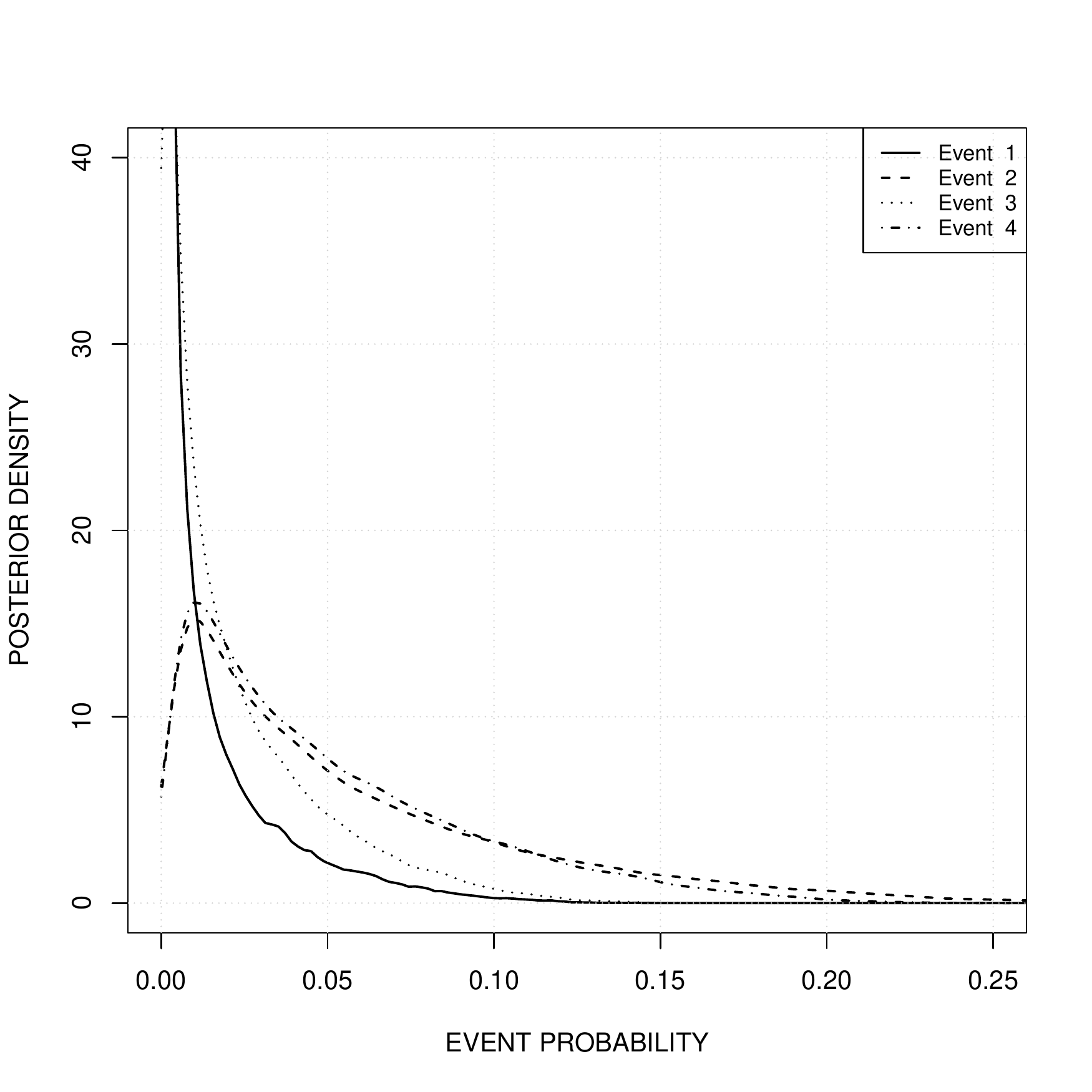}
  }
  \parbox{0.35\columnwidth}{
    \centering
    \includegraphics[scale=0.25]{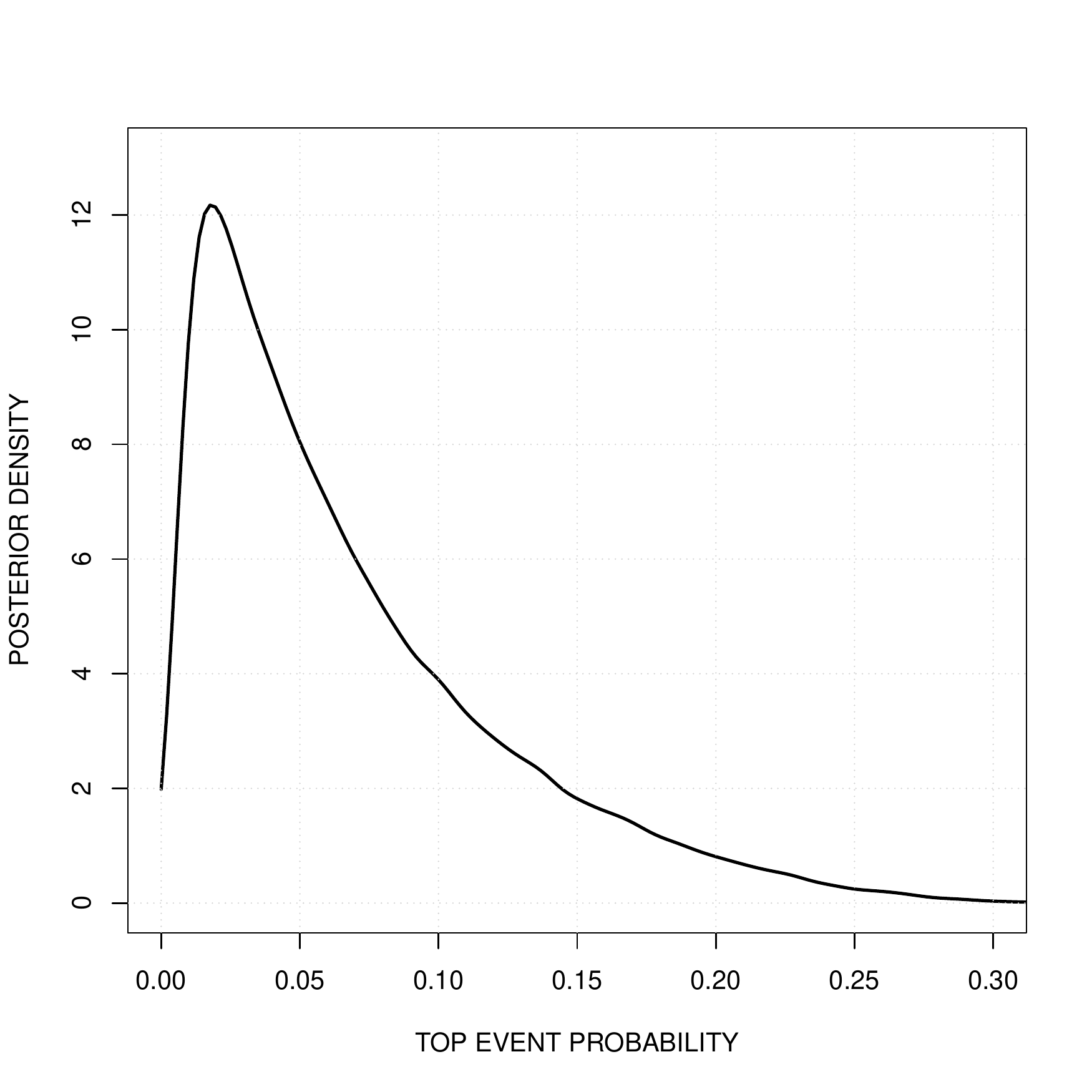}
}}
\parbox{\columnwidth}{  
  \parbox{0.3\columnwidth}{
     \textsf{\small{Randomly incomplete (50\%)\\
     $\mathbb{E}(\bm{p} \, | \, \mathcal{E}_{1:40}) \\ = (0.010,0.060,0.019,0.050)$ \\
     $\mathbb{E}(p_7 \, | \, \mathcal{E}_{1:40}) = 0.070$}}

  }   
  \parbox{0.35\columnwidth}{
    \centering
    \includegraphics[scale=0.25]{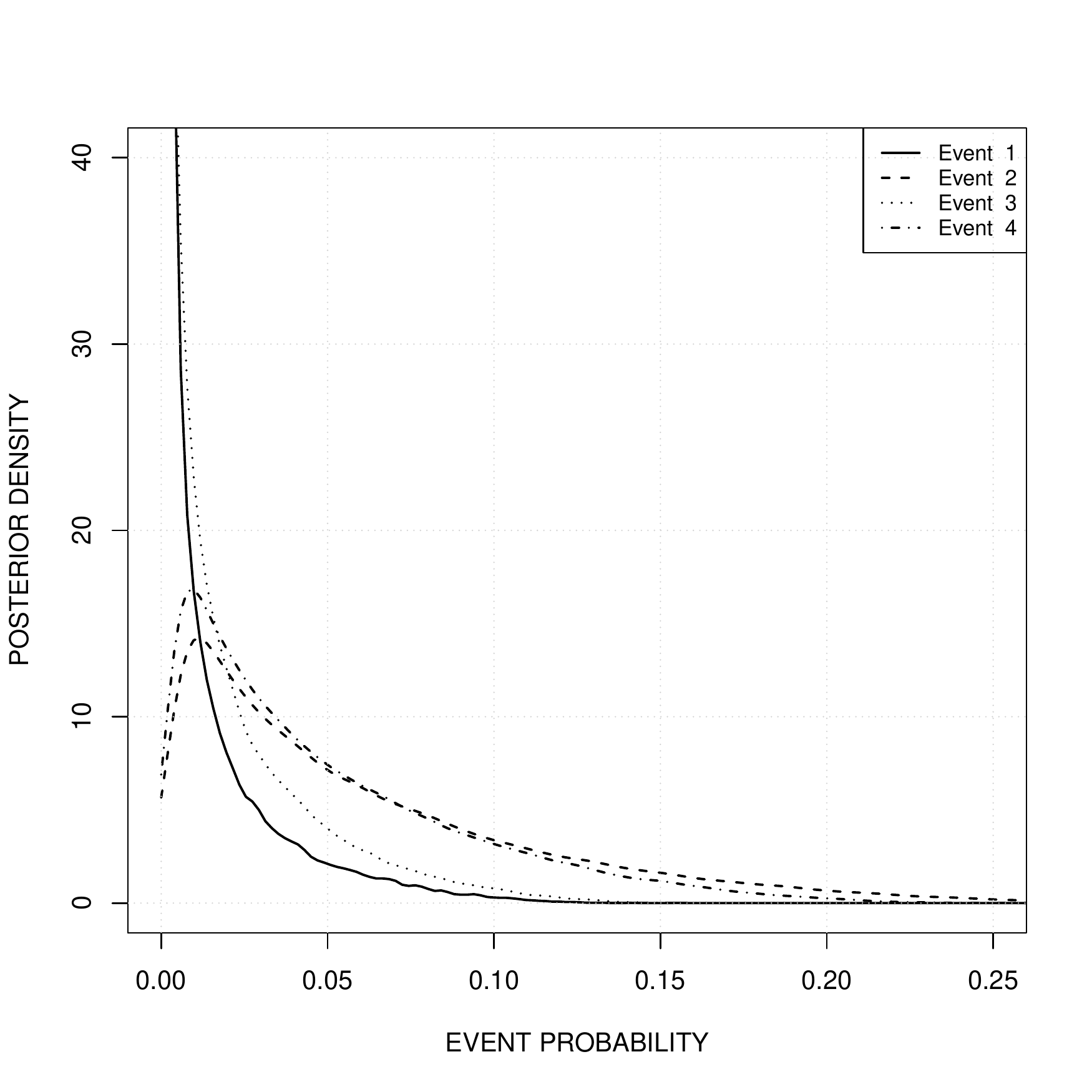}
  }
  \parbox{0.35\columnwidth}{
    \centering
    \includegraphics[scale=0.25]{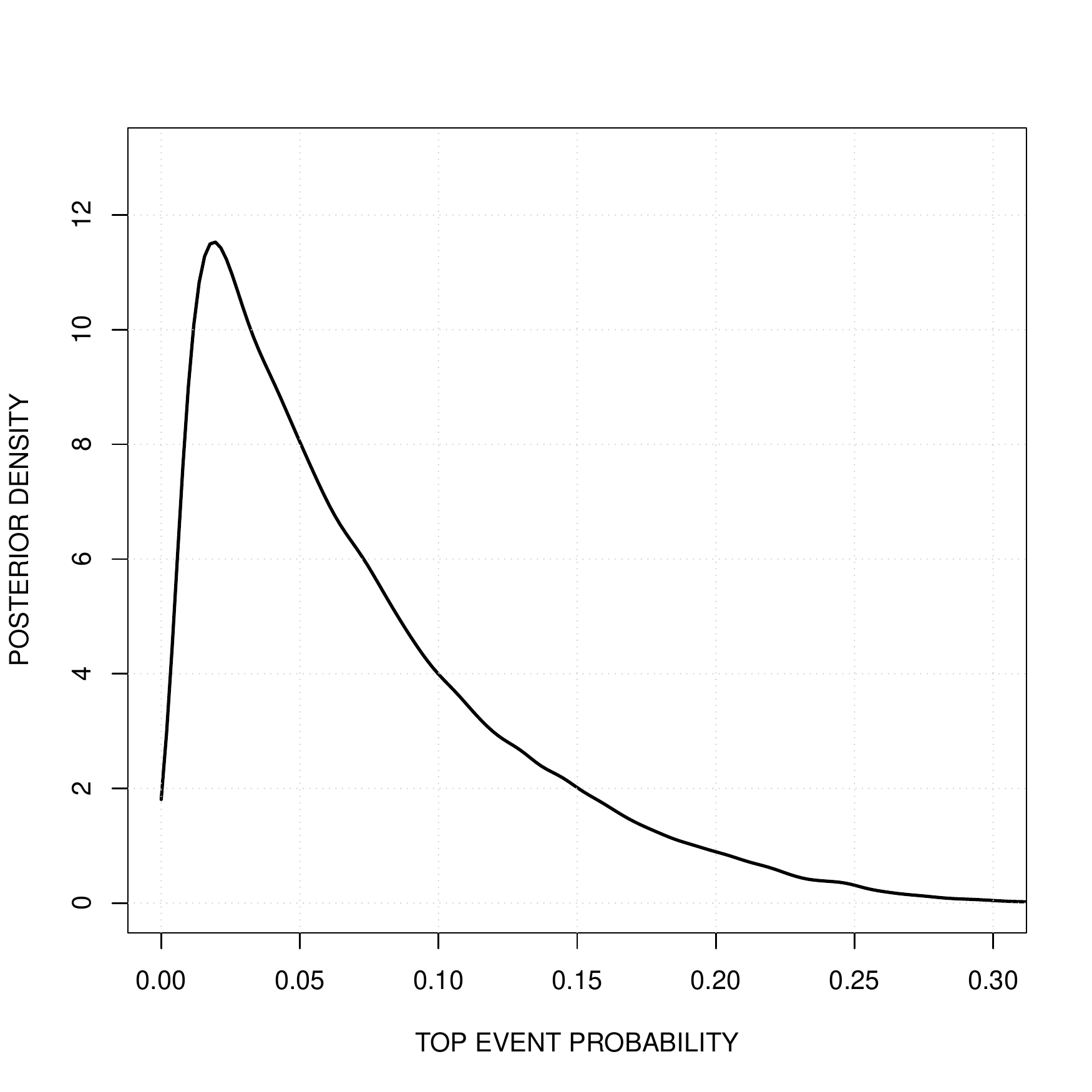}
}}
\parbox{\columnwidth}{  
  \parbox{0.3\columnwidth}{
     \textsf{\small{Top event only \\
     $\mathbb{E}(\bm{p} \, | \, \mathcal{E}_{1:40}) \\ = (0.012,0.054, 0.016, 0.044)$ \\
     $\mathbb{E}(p_7 \, | \, \mathcal{E}_{1:40}) = 0.065$}}
  }   
  \parbox{0.35\columnwidth}{
    \centering
    \includegraphics[scale=0.25]{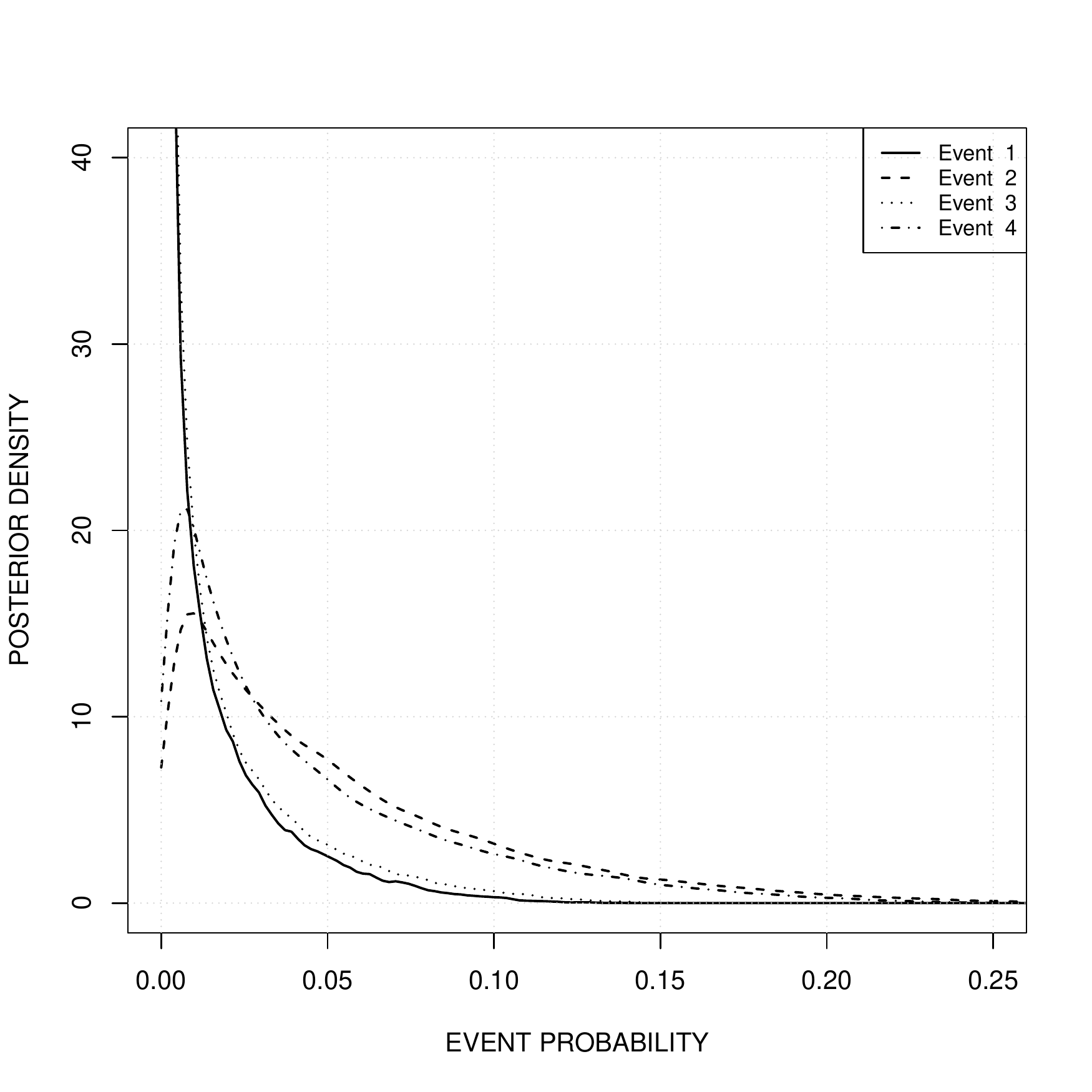}
  }
  \parbox{0.35\columnwidth}{
    \centering
    \includegraphics[scale=0.25]{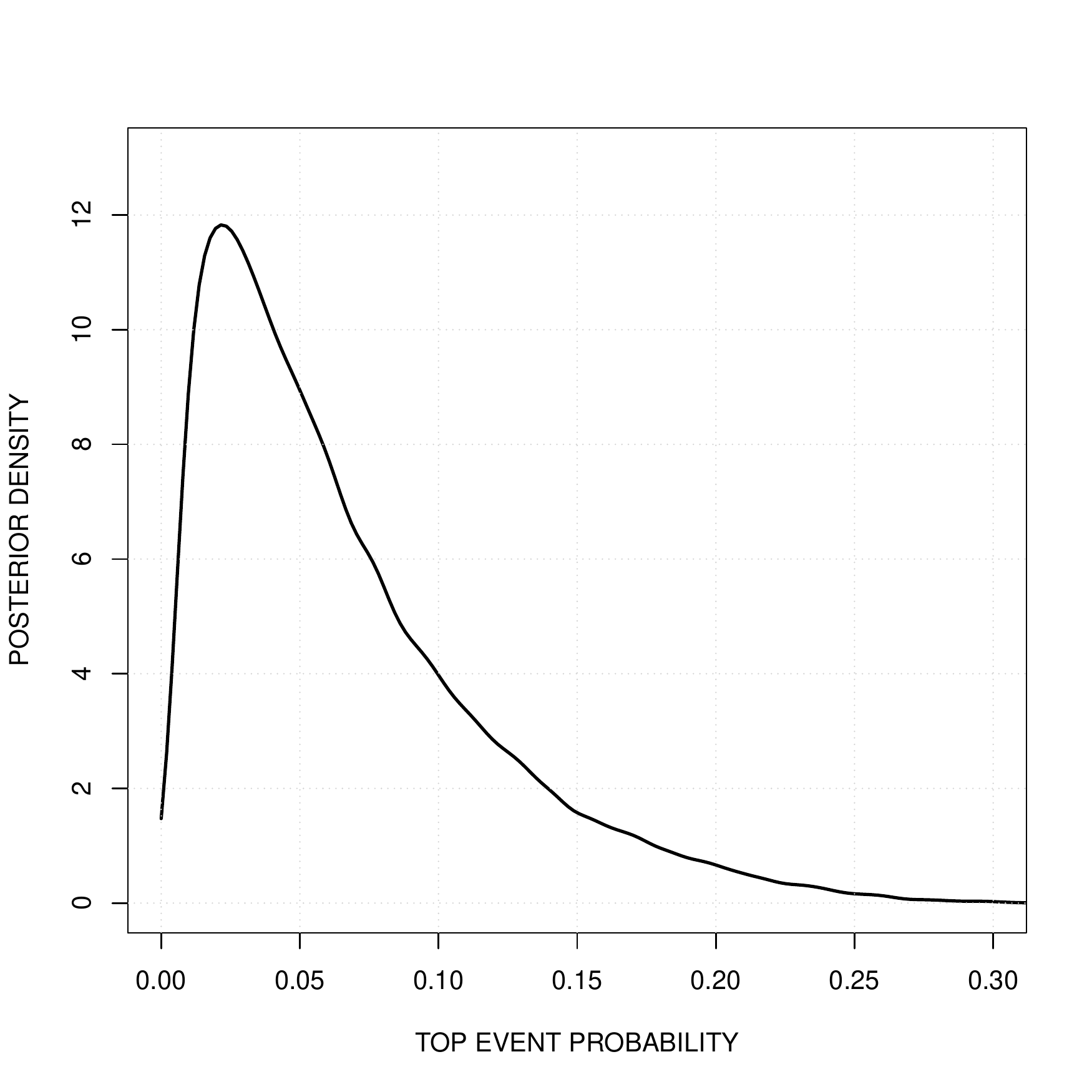}
}}
\caption{\label{fig:posterior_update_example} The prior and posterior distributions for the primary event and top event probabilities. The top row is the prior and subsequent rows are the posterior after 40 simulated observations of different types with $\bm{p} = (0.02,0.05,0.05,0.1)$ (and so a top event probability $p_7 = 0.0737$).}
\end{figure}

Figure \ref{fig:elicit_example_full} illustrated the prior elicitation stage for the simple system of Figure \ref{fig:simple_example}.  Figure \ref{fig:posterior_update_example} shows the result of posterior updating of the prior, as obtained from the incomplete set of comparisons in Figure \ref{fig:elicit_example_full},  after 40 randomly generated observations of this system with primary event probabilities $(0.02,0.05,0.05,0.10)$, giving a top event probability of $p_7 = 0.074$. This is done for 3 cases:
\begin{itemize}
\item The data set is complete, with the value of all events observed;
\item The same data set but incomplete, with each event randomly observed with probability 0.5;
\item The same data set but with the top event observed only;

\end{itemize}
Implementation was with the random walk Metropolis algorithm over 1,000,000 iterations with $s_{\lambda}^2 = 0.25$. The figure shows very little difference in the inference between complete and 50\% incomplete data.  This is not surprising in this case as, in many cases, it is possible to logically infer most if not all of the missing values.  Posterior uncertainty using the top event observation only is somewhat larger than in the other 2 cases.

To further explore this, each of these posterior updates was repeated 100 times, each time with data regenerated with the same true primary event probabilities. Table \ref{tab:simstudy_uniform} summarises the posterior distributions obtained over these 100 runs where a uniform prior has been assumed for the primary event probabilities, while Table \ref{tab:simstudy_incomplete} repeats this for the prior obtained from the incomplete comparisons in Figure \ref{fig:elicit_example_full}. Each table shows the average of the posterior means, standard deviations and central 95\% probability intervals, as well as the root mean square error between the posterior mean and the true value:
\[ \mbox{RMSE} \: = \: \left( \frac{1}{100}\sum_{r=1}^{100} (p_{\mbox{\scriptsize{posterior}},r} - p_{\mbox{\scriptsize{true}}})^2 \right)^{1/2}, \]
where $p_{\mbox{\scriptsize{posterior}},r}$ is the posterior mean of $p$ from the $r$th run and $p_{\mbox{\scriptsize{true}}}$ is the true value.

In general, the tables show that the data have the effect of moving prior distributions in the direction of the true values.  The prior coming from the elicitation is quite strong and has a large effect on the posterior, relative to the uniform prior case, even after 40 observations. For the primary event probabilities, the greater information in the complete data does this more than the incomplete, which does this more than the top event only data. However, note that inference for the top event probability is very similar across all 3 data types.

\begin{table}
\centering
\begin{tabular}{lccccc}
                              & $\bm{p_1}$ & $\bm{p_2}$ & $\bm{p_3}$ & $\bm{p_4}$ & $\bm{p_7}$     \\ \hline
\bf{True Value}                 & 0.020 & 0.050 & 0.050 & 0.100 & 0.074     \\ 
\bf{Prior Mean}                 & 0.500 & 0.500 & 0.500 & 0.500 & 0.812     \\ \hline
                                & \multicolumn{5}{c}{\it Top Event Observation Only} \\
\bf{Average of: }                  &  &  &  &  & \\
\hspace{2mm}post.\ means      & 0.029 & 0.037 & 0.160 & 0.146 & 0.073 \\
\hspace{2mm}post.\ std.\ devs & 0.027 & 0.032 & 0.258 & 0.258 & 0.038 \\
\hspace{2mm}post. 95\% PIs    & (0.001,0.098) & (0.003,0.115) & (0.0001,0.703) & (0.0001,0.715) & (0.019,0.165) \\

\bf{RMSE}                          & 0.0366 & 0.0414 & 0.1938 & 0.1445 & 0.0414 \\ 
\hline
 & \multicolumn{5}{c}{\it Randomly Incomplete (50\%)} \\
\bf{Average of:}                   &  &  &  &  & \\
\hspace{2mm}post.\ means      & 0.020 & 0.054 & 0.055 & 0.098 & 0.078 \\
\hspace{2mm}post.\ std.\ devs & 0.016 & 0.033 & 0.040 & 0.057 & 0.040 \\
\hspace{2mm}post. 95\% PIs    &  (0.002,0.057) & (0.011,0.133) & (0.007,0.153) & (0.020,0.236) & (0.021,0.173) \\
\bf{RMSE}                           & 0.0290 & 0.0415 & 0.0483 & 0.0640 & 0.0465 \\ \hline
                              & \multicolumn{5}{c}{\it Complete Observation} \\
\bf{Average of:}                   &  &  &  &  & \\
\hspace{2mm}post.\ means      & 0.021 & 0.053 & 0.057 & 0.103 & 0.078 \\
\hspace{2mm}post.\ std.\ devs & 0.017 & 0.032 & 0.033 & 0.045 & 0.038 \\
\hspace{2mm}post. 95\% PIs    &  (0.002,0.065) & (0.011,0.130) & (0.013,0.138) & (0.034,0.207) & (0.024,0.168) \\
\bf{RMSE}                          & 0.0221 & 0.0346 & 0.0389 & 0.0481 & 0.0428 \\ \hline
\end{tabular}
\caption{\label{tab:simstudy_uniform}Summary of results for the simulation study of the system in Figure \ref{fig:simple_example}, using a uniform prior on each primary event probability.}
\end{table}

\begin{table}
\centering
\begin{tabular}{lccccc}                                                      \hline
                              & $\bm{p_1}$ & $\bm{p_2}$ & $\bm{p_3}$ & $\bm{p_4}$ & $\bm{p_7}$  \\ \hline
\bf{True Value}                    & 0.020 & 0.050 & 0.050 & 0.100 & 0.074     \\ 
\bf{Prior Mean}                    & 0.020 & 0.077 & 0.023 & 0.054 & 0.097     \\ \hline
                              & \multicolumn{5}{c}{\it Top Event Observation Only} \\
\bf{Average of:}                   &  &  &  &  & \\
\hspace{2mm}post.\ means      & 0.013 & 0.059 & 0.016 & 0.045 & 0.071 \\
\hspace{2mm}post.\ std.\ devs & 0.020 & 0.051 & 0.023 & 0.044 & 0.052 \\
\hspace{2mm}post. 95\% PIs    & (0.000,0.072) & (0.003,0.191) & (0.000,0.084) & (0.002,0.163) & (0.009,0.204) \\

\bf{RMSE}                          & 0.0076 & 0.0138 & 0.0341 & 0.0553 & 0.0125 \\  \hline
                              & \multicolumn{5}{c}{\it Randomly Incomplete (50\%)} \\
\bf{Average of:}                   &  &  &  &  & \\
\hspace{2mm}post.\ means      & 0.013 & 0.058 & 0.019 & 0.050 & 0.071 \\
\hspace{2mm}post.\ std.\ devs & 0.020 & 0.053 & 0.024 & 0.044 & 0.055 \\
\hspace{2mm}post. 95\% PIs    & (0.000,0.072) & (0.003,0.198) & (0.0001,0.087) & (0.003,0.162) & (0.008,0.212) \\
\bf{RMSE}                     & 0.0082 & 0.0133 & 0.0315 & 0.0480 & 0.0119 \\ \hline
                              & \multicolumn{5}{c}{\it Complete Observation} \\
\bf{Average of:}                   &  &  &  &  & \\
\hspace{2mm}post.\ means      & 0.014   & 0.055  & 0.023  & 0.057  & 0.069  \\
\hspace{2mm}post.\ std.\ devs & 0.020   & 0.052  & 0.025  & 0.045  & 0.055  \\
\hspace{2mm}post. 95\% PIs    & (0.000,0.075) & (0.003,0.194) & (0.0003,0.091) & (0.004,0.166) & (0.008,0.210)\\
\bf{RMSE}                         &  0.0078 & 0.0123 & 0.0278 & 0.0448 & 0.0117 \\ \hline
\end{tabular}
\caption{\label{tab:simstudy_incomplete}Summary of results for the simulation study of the system in Figure \ref{fig:simple_example}, using the incomplete pairwise comparison prior of Figure \ref{fig:elicit_example_full}.}
\end{table}

\section{Application to ATV re-entry}
\label{sec:ATV}

The Autonomous Transfer Vehicle (ATV) was developed by the European Space Agency to re-supply the International Space Station.  Its mission consisted of a launch to the station with supplies (e.g. experimental equipments, propellants and goods for the permanent crew); once unloaded at the station, waste was placed into it.  The vehicle undocked from the station and was designed to have a controlled burn-up in the atmosphere, with any surviving fragments landing in the remote South Pacific. Five ATVs were launched between 2008 and 2014. All successfully resupplied the Space Station and were then successfully de-orbited.

One risk associated with the ATV, and most spacecraft that are de-orbited, is that the vehicle will unexpectedly explode during re-entry, due to causes such as the heating of leftover fuel, which could result in scattering debris over land \cite{koppenwallner05}. Due to the difficulties in observing such a re-entry, there is unlikely to be complete information on the causes of an explosion \cite{depasquale09}.  The question then arises as to what can be learned from the observation. \cite{depersis16} conducted a fault tree analysis as part of an assessment of the risk of this event, from which we base this analysis.

\begin{figure}
\centering
\includegraphics[scale=0.55]{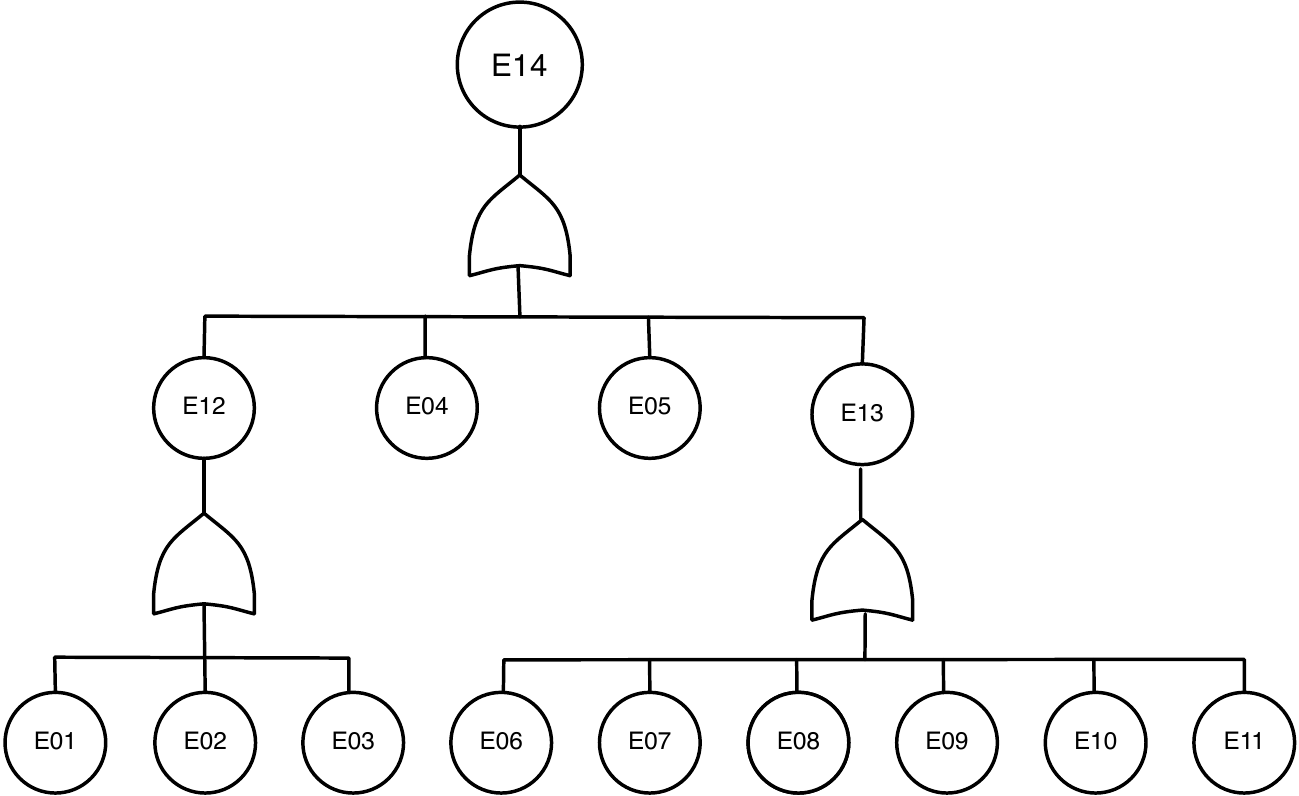}
\caption{\label{fig:ATV_tree} Fault tree for unexpected explosion during re-entry of the Autonomous Transfer Vehicle.}
\end{figure}

Figure \ref{fig:ATV_tree} shows the fault tree that was elicited from expert engineers at ESA. The Appendix gives a description of each node in the tree. Note that the tree contains only OR nodes, so that it only takes one of the primary events to occur in order for the top event to occur. Intermediate nodes are included in the diagram because they may be what is observable during the re-entry.

The primary nodes of the tree come in 3 groups:  
\begin{itemize}
\item[]Nodes 1 to 3 are events concerning propellant or the propellant tanks;
\item[]Nodes 6 to 11 are events concerning the batteries;
\item[]Nodes 4 and 5 are other causes of an explosion on re-entry.
\end{itemize}
Separate experts were consulted about each of these 3 groups.  Within each group, a complete pairwise comparison was made and the prior distributions were constructed separately from these complete comparisons within each group.  No comparisons between the 3 groups was made. The details of the elicitation are in the Appendix. Figure \ref{fig:ATV_priors} shows all the distinct prior distributions of primary events that were elicited and the implied prior on the top event probability. The prior mean for the top event probability is 0.17 with a central 95\% probability interval $(0.11,0.24)$.

\begin{figure}
\centering
\includegraphics[scale=0.4]{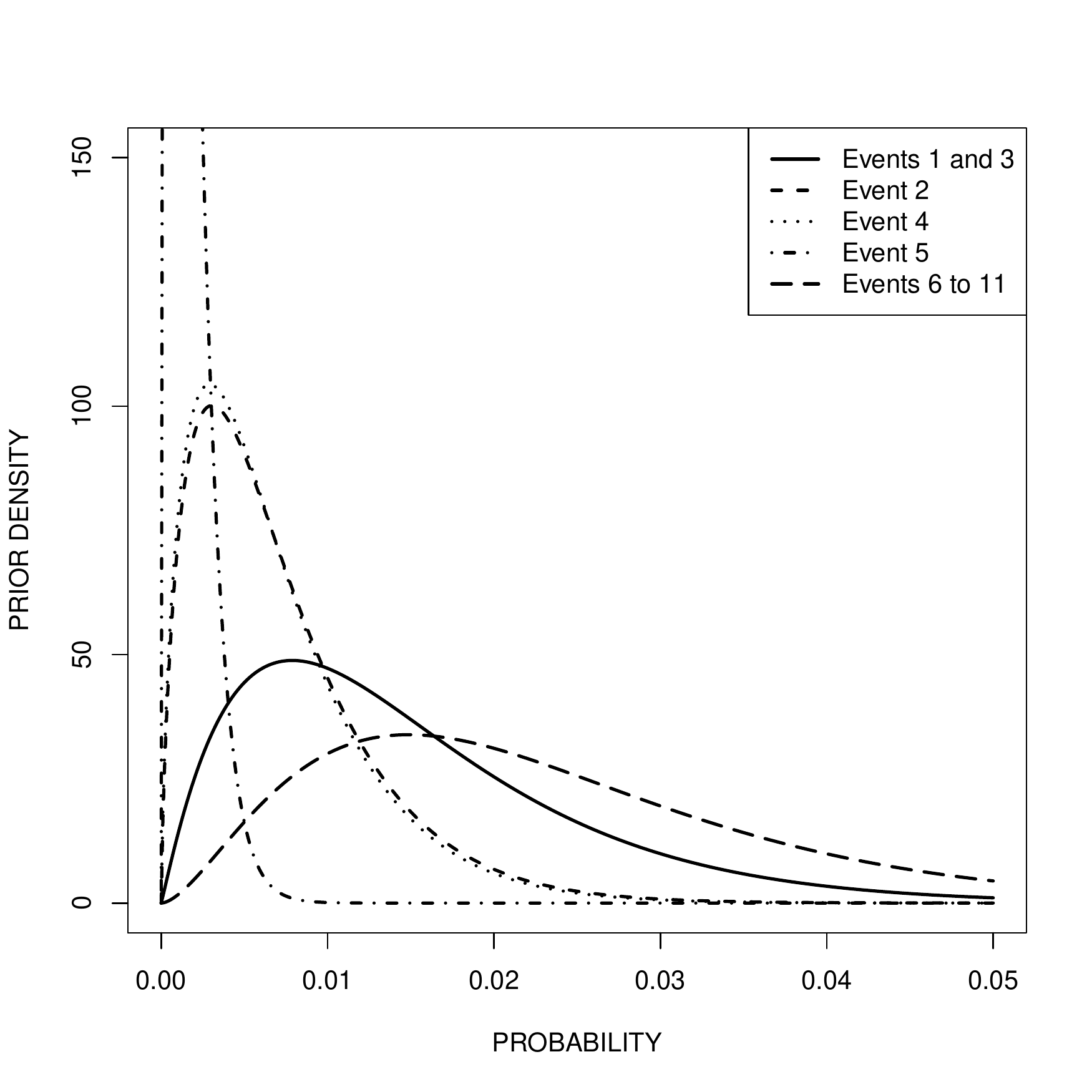}
\includegraphics[scale=0.4]{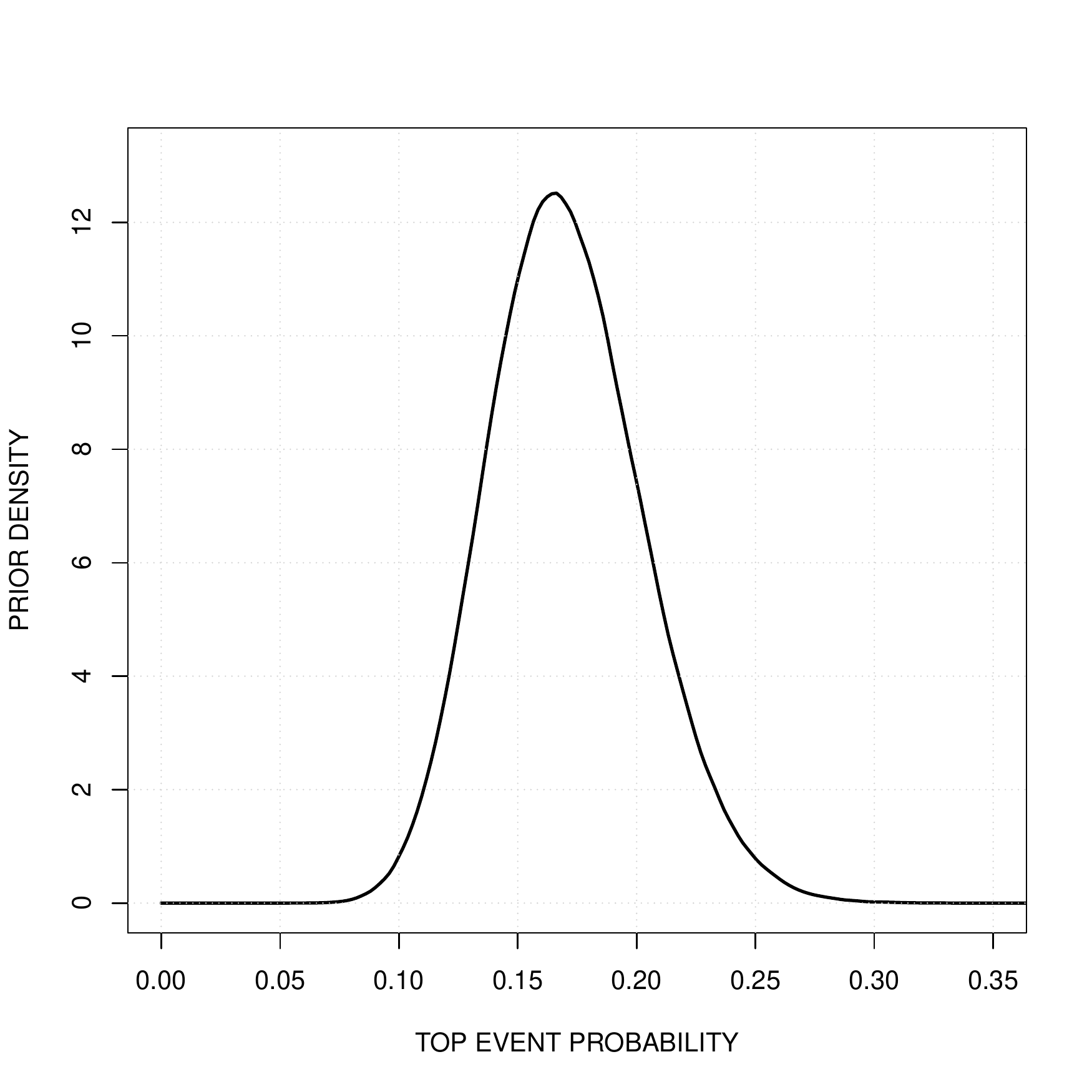} 
\caption{\label{fig:ATV_priors} Prior distributions on primary event probabilities (left) and the resulting prior on the top event probability (right), based on the elicitation described in the Appendix.}
\end{figure}

Figure \ref{fig:atv_posterior_topeventonly} shows a kernel density estimate of the posterior distribution of the top event probability, given observation that all 5 re-entries occurred without explosion.  These data are just observation of the top event; however, because all nodes in the fault tree are OR gates, observation that the top event did not occur is equivalent to a complete observation that all primary events did not occur.  As expected, the posterior distribution is shifted towards smaller probabilities.  The posterior mean probability is now 0.10, with a central 95\% probability interval of $(0.02,0.25)$. Compared to the prior, the mean has decreased but uncertainty in the value of the top event probability has actually increased because the data are somewhat in contradiction with the prior opinion. 
\begin{figure}
\centering
\includegraphics[scale=0.4]{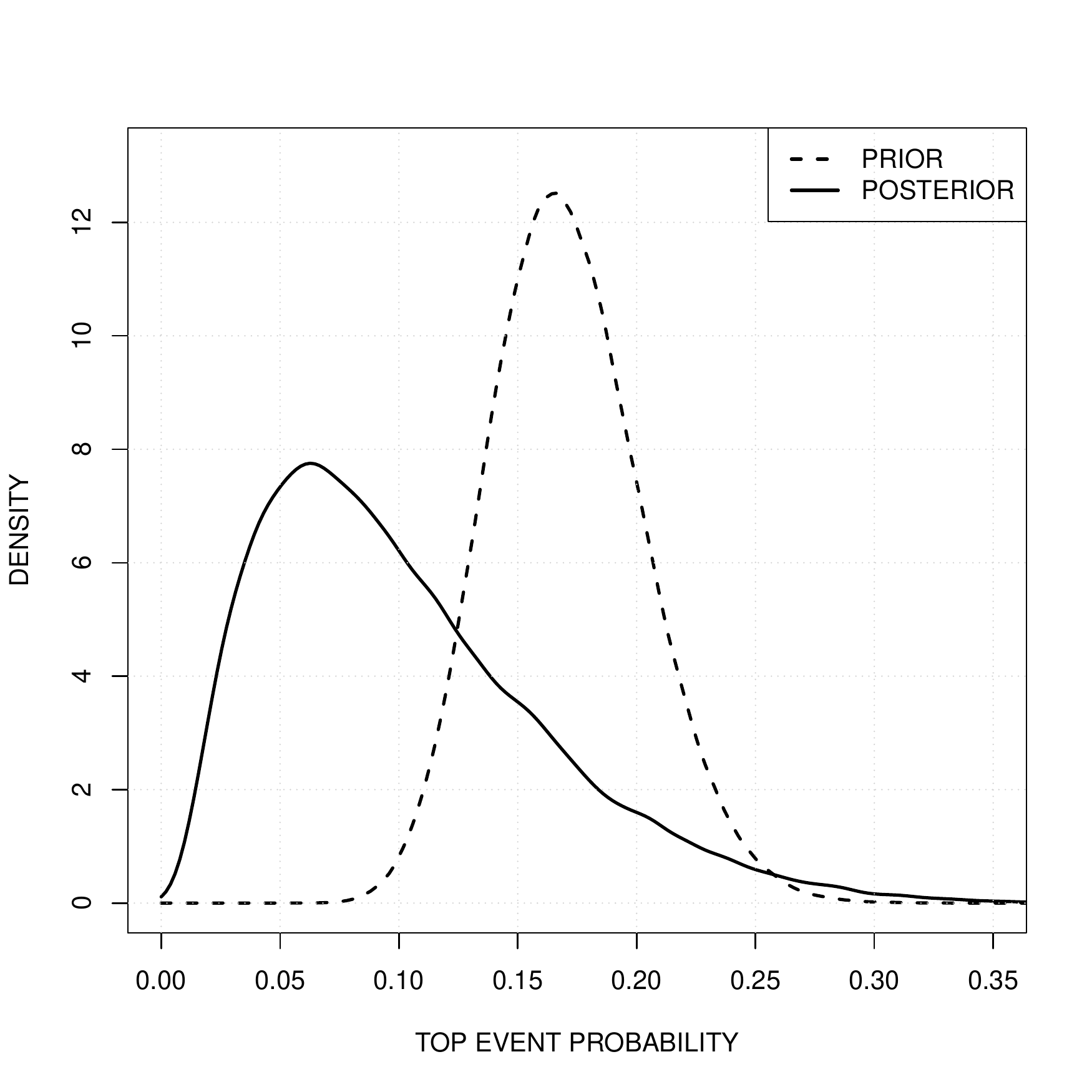}
\caption{\label{fig:atv_posterior_topeventonly} The posterior distribution of the top event probability given data that all 5 ATV spacecraft were observed to re-enter without explosion.  The prior from the elicitation in the Appendix is shown for comparison.}
\end{figure}

 A detailed observation campaign of the first ATV re-entry was attempted at considerable cost, using aircraft-based cameras that tracked the re-entry trajectory \cite{lips10}. No similar observation campaign was attempted subsequently.  In this section an exploration of the value of an observation campaign, that was able to observe some or all of the intermediate and primary events in the tree, is explored through simulation.  The events in the tree are simulated using the elicited prior means as the primary event probabilities. Table \ref{tab:observation_campaign_ATV} summarises the posterior uncertainty in the top event probability under different combinations of data type and size.  For data size, we look at one observation (being the number of observation campaigns actually undertaken), 5 (being the total number of ATVs that were de-orbited) and then 20 and 100, to see what is the impact of a campaign on a satellite series with more de-orbits. For data type, we consider 3 different observation campaigns: observation of the top event only, observation of the top and intermediate events only, and observation of all events in the tree. The table is a summary of the inference of each combination over 100 simulated data sets.
 
The principal feature of the table is that there appears to be very little difference in posterior uncertainty between the 3 types of observation.  This makes sense in this case since (a) the prior expectation is that about 83\% of top event observations will be that no explosion occurred and (b) such an observation is logically equivalent to the complete observation that all primary events did not occur.  In other words, in most cases there is no real difference between the different data types.  To emphasise this, Table \ref{tab:observation_campaign_ATV_uniform} repeats the analysis with uniform priors on all of the primary event probabilities, and a similar behaviour is seen.

The conclusion is that, under this fault tree, there is little benefit to a detailed observation campaign, at least in terms of quantifying the risk of the top event.  We note that there are other reasons to conduct an observation campaign outside the scope of this paper, such as characterisation of the fragment size during spacecraft break-up, the altitude and velocity of those fragments, etc.

\begin{table}
\centering
\begin{tabular}{lcccc}                                           \hline
\bf{Data Size}                   & \bf{1} & \bf{5} & \bf{20} & \bf{100} \\ \hline
\it{\bf{Prior}}                  &        &        &         &  \\
\hspace{2mm}Means (and true value) & 0.171 & 0.171 & 0.171   &  0.171 \\
\hspace{2mm}Std.\ devs           & 0.032  & 0.032  & 0.032   &  0.032 \\
\hspace{2mm}95\% PI width        & 0.125  & 0.125  & 0.125   &  0.125 \\ \hline                                
\it{\bf{Top Event}}              &        &        &         &        \\
\bf{Average of:}                 &        &        &         &        \\
\hspace{2mm}post.\ means         & 0.106  & 0.108  & 0.112   & 0.128  \\
\hspace{2mm}post.\ std.\ devs    & 0.061  & 0.060  & 0.058   & 0.048  \\
\hspace{2mm}post. 95\% PI widths & 0.228  & 0.229  & 0.221   & 0.192  \\ \hline
\it{\bf{Top \& Intermediate Events}} &    &        &         &        \\
\bf{Average of:}                 &        &        &         &        \\
\hspace{2mm}post.\ means         &  0.107 &  0.107 & 0.112   & 0.127  \\
\hspace{2mm}post.\ std.\ devs    &  0.061 &  0.060 & 0.058   & 0.049  \\
\hspace{2mm}post. 95\% PI widths &  0.230 &  0.228 & 0.221   & 0.195  \\ \hline
\it{\bf{Complete Observation}}   &        &        &         &        \\
\bf{Average of:}                 &        &        &         &        \\
\hspace{2mm}post.\ means         & 0.106  & 0.107 & 0.112 & 0.127 \\
\hspace{2mm}post.\ std.\ devs    & 0.061 & 0.060  & 0.059   & 0.057  \\
\hspace{2mm}post. 95\% PI widths & 0.229  & 0.230  & 0.226   & 0.218  \\ \hline
\end{tabular}
\caption{\label{tab:observation_campaign_ATV} Average posterior uncertainty in the top event probability for different combinations of data size and observation campaign, under the elicited prior.}
\end{table} 

\begin{table}
\centering
\begin{tabular}{lcccc}                                           \hline
\bf{Data Size}                   & \bf{1} & \bf{5} & \bf{20} & \bf{100} \\ \hline
\it{\bf{True Value}}             & 0.171  & 0.171  & 0.171   &  0.171   \\ \hline
\it{\bf{Prior}}                  &        &        &         &          \\
\hspace{2mm}Means                & 0.9995 & 0.9995 & 0.9995  & 0.9995   \\
\hspace{2mm}Std.\ devs           & 0.002  & 0.002  & 0.002   & 0.002    \\
\hspace{2mm}95\% PI width        & 0.004  & 0.004  & 0.004   & 0.004    \\ \hline                                
\it{\bf{Top Event}}              &        &        &         &          \\
\bf{Average of:}                 &        &        &         &          \\
\hspace{2mm}post.\ means         & 0.269  & 0.188  & 0.169   & 0.176    \\
\hspace{2mm}post.\ std.\ devs    & 0.164  & 0.128  & 0.078   & 0.038    \\
\hspace{2mm}post. 95\% PI widths & 0.644  & 0.462  & 0.297   & 0.146    \\ \hline
\it{\bf{Top \& Intermediate Events}} &    &        &         &          \\
\bf{Average of:}                 &        &        &         &          \\
\hspace{2mm}post.\ means         & 0.716  & 0.173  & 0.184   & 0.171    \\
\hspace{2mm}post.\ std.\ devs    & 0.372  & 0.113  & 0.080   & 0.037    \\
\hspace{2mm}post. 95\% PI widths & 0.925  & 0.428  & 0.306   & 0.143    \\ \hline
\it{\bf{Complete Observation}}   &        &        &         &          \\
\bf{Average of:}                 &        &        &         &          \\
\hspace{2mm}post.\ means         & 0.297  & 0.184  & 0.170   & 0.176    \\
\hspace{2mm}post.\ std.\ devs    & 0.171  & 0.127  & 0.075   & 0.036    \\
\hspace{2mm}post. 95\% PI widths & 0.673  & 0.459  & 0.284   & 0.140    \\ \hline
\end{tabular}
\caption{\label{tab:observation_campaign_ATV_uniform} Average posterior uncertainty in the top event probability for different combinations of data size and observation campaign, under a uniform prior.}
\end{table}

\section{Conclusion}
\label{sec:conc}
An approach to probabilistic risk assessment for a system, using a combination of fault tree analysis, prior elicitation and Bayesian updating, has been described. Through its use of pairwise comparisons for prior elicitation, it has particular use in circumstances where the access to experts is limited or they have little experience of elicitation methods.  The method is illustrated with an application to spacecraft re-entry risk. 

The principal benefits of the approach are the relatively light burden on the expert for elicitation, and the proper management of uncertainties through a Bayesian network.  Any other elicitation approach for the primary event probabilities can be 'plugged in' and used if needed, including more comprehensive methods that would give richer information.

\section*{Software}
Software, in the form of {\tt R} code, is available to implement the methods described in this paper.  At the time of writing, the code derives the likelihood using Equation \ref{eq:incomplete_likelihood} and implements inference by MCMC, as in Section \ref{subsec:compute}. The code includes sample scripts to run the analyses that generate Figures \ref{fig:elicit_example_full}, \ref{fig:posterior_update_example} and \ref{fig:atv_posterior_topeventonly}. The code is open source under GPL v3 license and can be downloaded from \url{https://github.com/jlbosque/RiskAssessment}.

\section*{Acknowledgements}
This research is partially supported by the NPI program of the European Space Agency and by the Spanish Ministry of Science, Innovation and Universities, under the program "Salvador de Madariaga", grant PRX18/00128. We are immensely grateful to Dr.\ Guillermo Ortega of ESA, who provided insight and expertise that greatly assisted the research.  We would like to thank those at the European Space Agency who gave their time and expertise to the elicitation process for the ATV application.

\bibliographystyle{chicago}
\bibliography{root}

\begin{thebibliography}{}

\bibitem[\protect\citeauthoryear{Aspinall and Cooke}{Aspinall and
  Cooke}{1997}]{aspinall13}
Aspinall, W. and R.~Cooke (1997).
\newblock Quantifying scientific uncertainty from expert judgement elicitation.
\newblock In J.~Rougier, S.~Sparks, and L.~J. Hill (Eds.), {\em Risk and
  Uncertainty Assessment for Natural Hazards}, pp.\  64--99. Cambridge
  University Press.

\bibitem[\protect\citeauthoryear{Barlow}{Barlow}{1988}]{barlow88}
Barlow, R.~E. (1988).
\newblock Using influence diagrams.
\newblock In C.~Clarotti and D.~Lindley (Eds.), {\em Accelerated Life Testing
  and Experts' Opinions in Reliability}, pp.\  145--157.

\bibitem[\protect\citeauthoryear{Bobbio, Portinale, Minichino, and
  Ciancamerla}{Bobbio et~al.}{2001}]{bobbio01}
Bobbio, A., L.~Portinale, M.~Minichino, and E.~Ciancamerla (2001).
\newblock Improving the analysis of dependable systems by mapping fault trees
  to {B}ayesian networks.
\newblock {\em Reliability Engineering \& System Safety\/}~{\em 71}, 249--260.

\bibitem[\protect\citeauthoryear{Bradley and Terry}{Bradley and
  Terry}{1952}]{bradley52}
Bradley, R.~A. and M.~Terry (1952).
\newblock Rank analysis of incomplete block designs {I}. the method of paired
  comparisons.
\newblock {\em Biometrika\/}~{\em 39}, 324--345.

\bibitem[\protect\citeauthoryear{Cagno, Caron, Mancini, and Ruggeri}{Cagno
  et~al.}{2000}]{cagno00}
Cagno, E., F.~Caron, M.~Mancini, and F.~Ruggeri (2000).
\newblock Using {A}{H}{P} in determining the prior distributions on gas
  pipeline failures in a robust {B}ayesian way.
\newblock {\em Reliability Engineering and System Safety\/}~{\em 67\/}(3),
  275--284.

\bibitem[\protect\citeauthoryear{Cooke}{Cooke}{1991}]{cooke91}
Cooke, R.~M. (1991).
\newblock {\em Experts in Uncertainty: Opinion and Subjective Probability in
  Science}.
\newblock Oxford: Oxford University Press.

\bibitem[\protect\citeauthoryear{Crawford and Williams}{Crawford and
  Williams}{1985}]{crawford85}
Crawford, G. and C.~Williams (1985).
\newblock A note on the analysis of subjective judgement matrices.
\newblock {\em Journal of Mathematical Psychology\/}~{\em 29\/}(4), 387--405.

\bibitem[\protect\citeauthoryear{de~Pasquale, Francillout, Wasbauer, Hatton,
  Albers, and Steele}{de~Pasquale et~al.}{2009}]{depasquale09}
de~Pasquale, E., L.~Francillout, J.~Wasbauer, J.~Hatton, J.~Albers, and
  D.~Steele (2009).
\newblock A{T}{V} {J}ules {V}erne reentry observation: {M}ission design and
  trajectory analysis.
\newblock In {\em IEEE Aerospace Conference}.

\bibitem[\protect\citeauthoryear{{De Persis}}{{De Persis}}{2016}]{depersis16}
{De Persis}, C. (2016).
\newblock {\em A risk assessment tool for highly energetic break-up events
  during the atmospheric re-entry}.
\newblock Ph.\ D. thesis, Trinity College Dublin.

\bibitem[\protect\citeauthoryear{Dias, Morton, and (editors)}{Dias
  et~al.}{2018}]{dias2018}
Dias, L.~C., A.~Morton, and J.~Q. (editors) (2018).
\newblock {\em Elicitation: the Science and Art of Structuring Judgement}.
\newblock Berlin: Springer.

\bibitem[\protect\citeauthoryear{Garthwaite, Kadane, and O'Hagan}{Garthwaite
  et~al.}{2005}]{garthwaite05}
Garthwaite, P.~H., J.~B. Kadane, and A.~O'Hagan (2005).
\newblock Statistical methods for eliciting probability distributions.
\newblock {\em J. Amer.\ Statist.\ Assoc.\/}~{\em 100}, 680--701.

\bibitem[\protect\citeauthoryear{Grimmett and Welsh}{Grimmett and
  Welsh}{2014}]{grimmett14}
Grimmett, G. and D.~Welsh (2014).
\newblock {\em Probability: an introduction\/} (Second ed.).
\newblock Oxford: Oxford University Press.

\bibitem[\protect\citeauthoryear{Gulliksen}{Gulliksen}{1959}]{gulliksen59}
Gulliksen, H. (1959).
\newblock Mathematical solutions for psychological problems.
\newblock {\em American Scientist\/}~{\em 47\/}(2), 178--201.

\bibitem[\protect\citeauthoryear{Guo and Sanner}{Guo and Sanner}{2010}]{guo10}
Guo, S. and S.~Sanner (2010).
\newblock Real-time multiattribute {B}ayesian preference elicitation with
  pairwise comparison queries.
\newblock In Y.~W. Teh and M.~Titterington (Eds.), {\em Proceedings of the 13th
  International Conference on Artificial Intelligence and Statistics
  (AISTATS)}, Volume~9 of {\em Proceedings of Machine Learning Research}, Chia
  Laguna Resort, Sardinia, Italy, pp.\  281--288. PMLR.

\bibitem[\protect\citeauthoryear{Koppenwallner, Fritsche, Lips, Martin,
  Francikkout, and {de Pasquale}}{Koppenwallner et~al.}{2005}]{koppenwallner05}
Koppenwallner, G., B.~Fritsche, T.~Lips, T.~Martin, L.~Francikkout, and E.~{de
  Pasquale} (2005).
\newblock Analysis of {A}{T}{V} destructive re-entry including explosive
  events.
\newblock In D.~Danesy (Ed.), {\em Proceedings of the 4th European Conference
  on Space Debris}, pp.\  545--550.

\bibitem[\protect\citeauthoryear{Langseth and Portinale}{Langseth and
  Portinale}{2007}]{langseth07}
Langseth, H. and L.~Portinale (2007).
\newblock Bayesian networks in reliability.
\newblock {\em Reliability Engineering \& System Safety\/}~{\em 92}, 92--108.

\bibitem[\protect\citeauthoryear{Lips, Lohle, Marynowsky, Rees,
  Stenbeak-Nielsen, Beks, and Hatton}{Lips et~al.}{2010}]{lips10}
Lips, T., S.~Lohle, T.~Marynowsky, D.~Rees, H.~Stenbeak-Nielsen, M.~Beks, and
  J.~Hatton (2010).
\newblock Assessment of the {A}{T}{V}-1 re-entry observation campaign for
  future re-entry missions.
\newblock In H.~Lacoste-Francis (Ed.), {\em Making Safety Matter: Proceedings
  of the Fourth IAASS Conference}.

\bibitem[\protect\citeauthoryear{Merrick and McLay}{Merrick and
  McLay}{2010}]{merrick10}
Merrick, J. R.~W. and L.~A. McLay (2010).
\newblock Is screening cargo containers for nuclear threats worthwhile?
\newblock {\em Decision Analysis\/}~{\em 7\/}(2), 155--171.

\bibitem[\protect\citeauthoryear{Mi, Li, Huang, Liu, and Zhang}{Mi
  et~al.}{2012}]{mi12}
Mi, J., Y.~Li, H.~Huang, Y.~Liu, and X.~Zhang (2012).
\newblock Reliability analysis of multi-state systems with common cause failure
  based on {B}ayesian networks.
\newblock In {\em Proceedings of the 2012 International Conference on Quality,
  Reliability, Risk, Maintenance, and Safety Engineering}, pp.\  1117--1121.

\bibitem[\protect\citeauthoryear{Neil, Taylor, and Marquez}{Neil
  et~al.}{2007}]{neil07}
Neil, M., M.~Taylor, and D.~Marquez (2007).
\newblock Inference in hybrid {B}ayesian networks using dynamic discretization.
\newblock {\em Statistics and Computing\/}~{\em 17}, 219--233.

\bibitem[\protect\citeauthoryear{O'Hagan, Buck, Daneshkhah, Eiser, Garthwaite,
  Jenkinson, Oakley, and Rakow}{O'Hagan et~al.}{2006}]{ohagan06}
O'Hagan, A., C.~E. Buck, A.~Daneshkhah, J.~R. Eiser, P.~H. Garthwaite, D.~J.
  Jenkinson, J.~E. Oakley, and T.~Rakow (2006).
\newblock {\em Uncertain judgements: eliciting experts' probabilities}.
\newblock Chichester: John Wiley \& Sons.

\bibitem[\protect\citeauthoryear{Saaty}{Saaty}{1980}]{saaty80}
Saaty, T.~L. (1980).
\newblock {\em The {A}nalytic {H}ierarchy {P}rocess: planning, priority
  setting, resource allocation}.
\newblock New York: McGraw-Hill.

\bibitem[\protect\citeauthoryear{Szwed, {van Dorp}, Merrick, Mazzuchi, and
  Singh}{Szwed et~al.}{2006}]{szwed06}
Szwed, P., J.~R. {van Dorp}, J.~R.~W. Merrick, T.~A. Mazzuchi, and A.~Singh
  (2006).
\newblock A {B}ayesian paired comparison approach for relative accident
  probability assessment with covariate information.
\newblock {\em European Journal of Operational Research\/}~{\em 169}, 157--177.

\bibitem[\protect\citeauthoryear{{Torres Toledano} and Sucar}{{Torres Toledano}
  and Sucar}{1998}]{torres98}
{Torres Toledano}, J. and L.~Sucar (1998).
\newblock Bayesian networks for reliability analysis of complex systems.
\newblock In {\em Lecture Notes in Artificial Intelligence}, Volume 1484, pp.\
  195--206. Berlin: Springer.

\bibitem[\protect\citeauthoryear{{van Dorp} and Merrick}{{van Dorp} and
  Merrick}{2010}]{vandorp10}
{van Dorp}, J.~R. and J.~R.~W. Merrick (2010).
\newblock On a risk management analysis of oil spill risk using maritime
  transportation system simulation.
\newblock {\em Annals of Operations Research\/}~{\em 187\/}(1), 249--277.

\bibitem[\protect\citeauthoryear{Verma, Ajit, and Muruva}{Verma
  et~al.}{2015}]{verma15}
Verma, A., S.~Ajit, and S.~. H.~P. Muruva (2015).
\newblock {\em Risk Management of Non-Renewable Energy Systems}.
\newblock New York: Springer.

\bibitem[\protect\citeauthoryear{Zhang and Thai}{Zhang and
  Thai}{2016}]{zhang16}
Zhang, G. and V.~V. Thai (2016).
\newblock Expert elicitation and {B}ayesian network modeling for shipping
  accidents: a literature review.
\newblock {\em Safety Science\/}~{\em 87}, 53--62.

\end{thebibliography}

\appendix
\noindent {\bf Derivation of the Prior Weights $\bm{w_i}$ from Pairwise Comparisons} \\
Our approach follows that of \cite{crawford85}, whose derivation of subjective weights from pairwise comparison scores has some advantages from a statistical point of view over the original eigenvector approach due to \cite{gulliksen59} and refined for AHP in \cite{saaty80}.  Let $q_{ij}$ be the comparison score (on the scale given in Figure \ref{fig:ahp_scores} in our case) between events $E_i$ and $E_j$. 

The weights are defined as geometric means of the pairwise comparison scores. In the case of elicitation of all pairwise comparisons, these are:
\[ w_i \: = \: \left( \prod_{\substack{j=1\\j \neq i}}^n q_{ij} \right)^{1/(n-1)}. \]

If only some of the pairwise comparisons have been made then the geometric mean is
\[ w_i \: = \: \left( \prod_{\substack{j=1\\j \neq i\\q_{ij} \mbox{\scriptsize{ defined}}}}^{n_i} q_{ij} \right)^{1/n_i}, \]
where $n_i$ is the number of comparisons made between $E_i$ and other events.

\vspace{3mm}

\noindent {\bf Derivation of $\bm{P(E_i \, | \, H_i \cap {\mathcal E}, \bm{p})}$} \\
Recall that $H_i$ are the ancestor events of $E_i$ and ${\mathcal E}$ are the observed events, so that $H_i \cap {\mathcal E}$ is the set of observed ancestors of $E_i$. In this section, this probability is derived in the case when $H_i \cap {\mathcal E}$ does not logically imply $E_i$, for which $P(E_i \, | \, H_i \cap {\mathcal E}, \bm{p}) = 1$. The possible logical implications are: $E_i=0 \Leftrightarrow \exists E_j \in \eta_i, E_j=1$ when $E_i$ is the result of an AND gate, or $E_i =1 \Leftrightarrow \exists E_j \in \eta_i, E_j=1$ when $E_i$ is the result of an OR gate. By working up the network from the observed primary events, any such implications are easy to determine.

If $E_i$ is a primary event then $H_i = \emptyset$ and $P(E_i \, | \, H_i \cap {\mathcal E}, \bm{p}) = p_i$.  If $E_i$ is not a primary event then we can write it in terms of its parents $\eta_i$ as 
\[ E_i \: = \: 
\begin{cases} 
\prod_{E_j \in \eta_i} E_j,           & \mbox{if AND gate,} \\
1 - \prod_{E_j \in \eta_i} (1 - E_j), & \mbox{if OR  gate.}
\end{cases} \]
If any events in $\eta_i$ are in $H_i \cap {\mathcal E}$ then their value is known and the above expressions for  $E_i$ conditional on $H_i \cap {\mathcal E}$ are
\begin{equation}
E_i \: = \: 
\begin{cases} 
\prod_{E_j \in \eta_i, \: E_j \notin {\mathcal E}}     E_j,           & \mbox{if AND gate,} \\
1 - \prod_{E_j \in \eta_i, \: E_j \notin {\mathcal E}} (1 - E_j),     & \mbox{if  OR  gate.}
\end{cases}
\label{eq:recursion_notobserved}
\end{equation}
Recursive application of Equation \ref{eq:recursion_notobserved} gives $E_i$ as a function of the unobserved primary events that are ancestors of $E_i$, which we denote $g_i$:
\begin{equation*}
P(E_i = e\, | \, H_i \cap {\mathcal E}, \bm{p}) \:
 = \: P(g_i(\{ E_j \, | \, j=1,\ldots,k; E_j \in H_i; E_j \notin {\mathcal E}\} = e \, | \, \bm{p});
 \end{equation*}
this function is a sum of products of $E_j$ and $(1-E_j)$ terms.   Since each $E_j$ is binary and independent given $\bm{p}$, the probability of this function is $g_i$ with each $E_j$ replaced by $p_j$:
\begin{equation*} 
P(E_i = e\, | \, H_i \cap {\mathcal E}, \bm{p}) \:
 = \: g_i(\{ p_j \, | \, j=1,\ldots,k; E_j \in H_i; E_j \notin {\mathcal E}\}).
 \end{equation*}

\vspace{3mm}

\noindent {\bf Definition of Node Events for the ATV Fault Tree} \\
Tables \ref{tab:ATV_primary_events} and \ref{tab:ATV_non_primary_events} give a description of the primary and non-primary events of the ATV fault tree in Figure \ref{fig:ATV_tree}.

\begin{table}[h]
\centering
\begin{tabular}{cp{45mm}} \hline
Event & Description \\ \hline \hline
E01 & Propellant valve leakage \\
E02 & Propellant tank destruction \\
E03 & Propellant pipe rupture\\ \hline
E04 & Pressure vessel burst, with sudden release of propellant  \\
E05 & Chemical reaction between hypergolic propellants\\ \hline
E06 & Battery over-pressure\\
E07 & Battery short circuit\\
E08 & Battery corrosion \\
E09 & Battery over-discharge\\
E10 & Battery over-temperature\\
E11 & Battery cell degradation\\ \hline \hline
\end{tabular}
\caption{\label{tab:ATV_primary_events}  Description of the primary events of the ATV fault tree in Figure \ref{fig:ATV_tree}.}
\end{table}

\begin{table}[h]
\centering
\begin{tabular}{cp{45mm}} \hline
Event & Description \\ \hline \hline
E12 & Chemical reaction of propellant and air \\
E13 & Burst of a battery cell \\
E14 & Top event. Explosion of spacecraft \\ \hline \hline
\end{tabular}
\caption{\label{tab:ATV_non_primary_events}  Description of the intermediate and top events of the ATV fault tree in Figure \ref{fig:ATV_tree}.}
\end{table}

\vspace{3mm}

\noindent {\bf Prior Elicitation for the ATV} \\
The pairwise comparison matrix for events E01, E02 and E03 was elicited as
\[ {\cal Q}_1 \: = \: \left( \begin{array}{ccc} 1.00 & 1.04 & 1.00 \\ 0.53 & 1.00 & 0.53 \\ 1.00 & 1.04 & 1.00 \end{array} \right). \]
The cornerstone event for this set of events is E01 and its interval was elicited to be $(0.01,0.04)$

The pairwise comparison matrix for events E04 and E05 was elicited as
\[ {\cal Q}_2 \: = \: \left( \begin{array}{cc} 1.00 & 1.23  \\ 0.28 & 1.00 \end{array} \right). \]
The cornerstone event for this set of events is E04 and its interval was elicited to be $(0.005,0.02)$

The pairwise comparison for events E06 to E11 was elicited as
\[ {\cal Q}_3 \: = \: \left( \begin{array}{cccccc} 1.00 & 1.00 & 1.00 & 1.00 & 1.00 & 1.00 \\ 1.00 & 1.00 & 1.00 & 1.00 & 1.00 & 1.00 \\ 1.00 & 1.00 & 1.00 & 1.00 & 1.00 & 1.00 \\ 1.00 & 1.00 & 1.00 & 1.00 & 1.00 & 1.00 \\ 1.00 & 1.00 & 1.00 & 1.00 & 1.00 & 1.00 \\ 1.00 & 1.00 & 1.00 & 1.00 & 1.00 & 1.00 \end{array} \right). \]
In this case, the expert evaluated that each of the primary events that concern a battery were equally likely.  The cornerstone event for this set of events is E06 and its interval was elicited to be $(0.014,0.055)$.

The resulting beta parameter values for each event probability prior are given in Table \ref{tab:ATV_prior}.

\begin{table}[t]
\centering
\begin{tabular}{lcccc} \hline
Event  & Prior         & Beta prior & \multicolumn{2}{c}{Prior}  \\ 
       & weights $w_i$ & parameters & Mean  & Mode \\ \hline \hline
E01, E03   & 0.400 & 2.06, 134.4  & 0.015  & 0.008  \\
E02        & 0.200 & 1.73, 237.7  & 0.007  & 0.003  \\ \hline
E04        & 0.815 & 1.72, 246.2  & 0.007  & 0.003  \\ 
E05        & 0.185 & 1.65, 1078.5 & 0.0015 & 0.0006 \\ \hline
E06 -- E11 & 0.167 & 2.70, 113.7  & 0.023  & 0.015  \\ \hline
\end{tabular}
\caption{\label{tab:ATV_prior} Prior elicitation results for the primary events of the ATV fault tree in Figure \ref{fig:ATV_tree}.}
\end{table}

\end{document}